\documentclass[11pt]{article}
\pdfoutput=1
\pdfoutput=1
\usepackage{geometry}                % See geometry.pdf to learn the layout options. There are lots.
\usepackage{cite}
\usepackage{graphicx}
\usepackage{amssymb}
\usepackage{mathtools}
\usepackage{epstopdf}
\usepackage{verbatim}
\usepackage{color}
\usepackage{slashed}
\usepackage{bbold}
\usepackage{fullpage}
\usepackage{float}
\usepackage[title]{appendix}
\usepackage[export]{adjustbox}
\usepackage[colorlinks, citecolor=blue, linkcolor=black]{hyperref}
\usepackage{amsfonts}
\usepackage{amssymb}
\usepackage{amsmath}
\usepackage{setspace}
\usepackage{epsfig}
\usepackage[font=footnotesize,labelfont=bf]{caption}
\usepackage[caption=false]{subfig}

\title{Realistic Dark Matter Candidate(s) from String/M-theory Hidden Sectors}

\newcommand{\Lagr}{\mathcal{L}}
\newcommand{\beq}{\begin{equation}}
\newcommand{\eeq}{\end{equation}}
\newcommand{\bmat}{\begin{pmatrix}}
\newcommand{\emat}{\end{pmatrix}}
\newcommand{\bal}{\begin{align}}
\newcommand{\eal}{\end{align}}

\newcommand{\Order}{\mathcal{O}}

\newcommand{\sigv}{\langle \sigma v \rangle}

\begin{document}

\def\simgt{\mathrel{\lower2.5pt\vbox{\lineskip=0pt\baselineskip=0pt
           \hbox{$>$}\hbox{$\sim$}}}}
\def\simlt{\mathrel{\lower2.5pt\vbox{\lineskip=0pt\baselineskip=0pt
           \hbox{$<$}\hbox{$\sim$}}}}

\begin{titlepage}
\begin{flushright}\small{MCTP-17-10} 
\end{flushright}

\begin{center}

{\LARGE \bf Categorisation and Detection of Dark Matter Candidates from String/M-theory Hidden Sectors}

\vspace{0.8cm}

\small
{\bf Bobby~S.~Acharya$^{1,2}$, Sebastian~A.~R.~Ellis$^{3}$, Gordon L. Kane$^{3}$,\\ Brent D. Nelson$^{4}$} and {\bf Malcolm Perry$^{5}$}
\normalsize

\vspace{.5cm}
{\it $^1$ Department of Physics, King's College London, London, WC2R 2LS, UK} \\
{\it $^2$ The Abdus Salam International Centre for Theoretical Physics, Strada Costiera 11, Trieste, Italy}\\
{\it $^3$ Michigan Center for Theoretical Physics, University of Michigan, Ann Arbor, MI 48109} \\
{\it $^4$ Department of Physics, Northeastern University, Boston, MA 02115, USA}\\
{\it $^5$ DAMTP, Centre for Mathematical Sciences, Wilberforce Road, Cambridge CB3 0WA, UK}

\abstract{\large{We study well-motivated dark matter candidates arising from weakly-coupled hidden sectors in compactified string/$M$-theory. Imposing generic top-down constraints greatly restricts allowed candidates. By considering the possible mechanisms for achieving the correct dark matter relic density, we compile categories of viable dark matter candidates and annihilation mediators. We consider the case where supersymmetry breaking occurs via moduli stabilisation and is gravitationally mediated to the visible and other hidden sectors, without assuming sequestering of the sector in which supersymmetry is broken. We find that in this case, weakly-coupled hidden sectors only allow for fermionic dark matter. Additionally, most of the mechanisms for obtaining the full relic density only allow for a gauge boson mediator, such as a dark $Z'$. Given these considerations, we study the potential for discovering or constraining the allowed parameter space given current and future direct detection experiments, and direct production at the LHC. We also present a model of a hidden sector which would contain a satisfactory dark matter candidate.}
}
\end{center}

\end{titlepage}
\tableofcontents
\section{Introduction}

Realistic string/$M$-theory compactifications typically predict an abundance of new states beyond the Minimal Supersymmetric Standard Model (MSSM). Generically, many of these states will be Standard Model (SM) gauge singlets, only interacting with MSSM fields via suppressed portal couplings and gravity. These ``hidden sector" particles are weakly constrained by energy and precision frontier experiments, in comparison to ``visible sector" particles with sizeable interactions with the SM. 

Despite the weak couplings between hidden and visible sector fields, the presence of hidden sectors in compactified string/$M$-theories has significant implications for early universe cosmology. This connection necessarily arises due to the presence of gravitationally coupled scalar fields known as moduli, whose expectation values determine the geometry of the compactified dimensions. The presence of moduli results in an epoch where the energy density of the universe may be dominated by coherent moduli oscillations. This moduli-dominated phase begins after inflation, and ends when the moduli decay to thermal bath particles, resulting in a radiation dominated universe shortly before big bang nucleosynthesis \cite{Preskill:1982cy, Abbott:1982af, Dine:1982ah, Pagels:1981ke, Weinberg:1982zq, Coughlan:1983ci, deCarlos:1993wie}. 

Canonical studies of stringy cosmological histories assume that the moduli decay exclusively to visible sector particles (SM particles and their superpartners). This is quite non-generic in the following sense. Since moduli only interact gravitationally, one generically expects moduli branching ratios to visible and hidden sector particles to be comparable. Thus hidden sector particles should be produced at the end of modulus domination with abundances similar to or larger than those of visible sector particles. If any hidden sector particles are stable due to symmetry or kinematic reasons, they will give an irreducible contribution to the relic dark matter (DM) abundance. These effects must be accounted for to give a complete picture of the early universe cosmology generically predicted by compactified string/M-theories. Hidden sector particles might also affect the expansion rate of the early universe. 

Hidden sector particles might indeed contribute all of the relic DM abundance. Recently, it has been argued that generic string/$M$-theory compactifications imply the lightest supersymmetric particle in the visible sector (LVSP) is unstable~\cite{Acharya:2016fge}. This is the result of string/$M$-theory compactifications generally having the ingredients necessary for the LVSP to decay, namely: 1) the existence of at least one hidden sector; 2) a portal connecting the visible sector to the hidden sector; 3) the hidden sector spectrum containing a suitable particle lighter than the LVSP. These conditions, if met, will result in the visible sector not containing a suitable dark matter candidate, further motivating the need for studying hidden sector dark matter candidates. It should be emphasised that the LVSP will decay if any hidden sector satisfies these conditions, even if many do not.

In addition to the ``top-down" motivations outlined above, there are also phenomenological motivations for considering the cosmology of stringy hidden sectors. 
Consider the lightest supersymmetric particle in the MSSM (the MSSM-LSP). This is the simplest candidate for the LVSP in a string-motivated model.
It is well known that for the moduli-dominated cosmologies mentioned above, while a pure Bino- or Higgsino-LSP is typically overabundant, an $\mathcal{O}(100)$ GeV Wino-LSP seems to provide a viable DM candidate~\cite{Moroi:1999zb,Acharya:2008bk}. However, a stable $\mathcal{O}(100)$ GeV Wino is in tension with recent indirect detection constraints~\cite{Cohen:2013ama,Fan:2013faa} unless a hierarchy is present between the moduli and gravitino masses~\cite{Fan:2013faa}. This conclusion was generalized in~\cite{Blinov:2014nla} to any stable MSSM-LSP. A simple way to avoid these constraints without imposing an unnatural moduli-gravitino mass hierarchy (see e.g.~\cite{Denef:2004cf,Acharya:2010af}) is for the visible sector LSP to decay to hidden sector particle(s), a possibility which is well-motivated from the ``top-down" perspective~\cite{Acharya:2016fge}, as described above. A corollary is that hidden sector particles can make up a sizeable portion (perhaps all) of the observed DM relic abundance, without violating any existing experimental constraints. Indeed, if one thinks seriously about the presence of hidden sectors with stable particles, it seems generically likely that an $\mathcal{O}(100)$ GeV LSP in the visible sector will decay to lighter hidden sector particles via portal couplings (e.g. kinetic mixing~\cite{Holdom:1985ag,Dienes:1996zr}). 

In this work, we systematically study the cosmological implications of generic, weakly-coupled, hidden sectors predicted in compactified string/$M$-theories.\footnote{For recent studies of strongly-coupled hidden sectors motivated from string theory, see~\cite{Halverson:2016nfq,Acharya:2017szw,Dienes:2016vei,Soni:2016gzf,Soni:2017nlm}.} This may may seem like a daunting task, due to the wide variety of hidden sectors which can plausibly arise in compactified string/$M$-theories. However, one can make considerable progress by combining general ``top-down" reasoning with cosmological/phenomenological constraints to determine generic criteria for viable string-motivated hidden sectors. This will considerably narrow the viable possibilities, allowing for a more tractable categorisation of realistic string-motivated hidden sectors. As part of our analysis, we also elucidate the conditions required to allow the MSSM-LSP to decay to hidden sector particles and avoid the indirect detection constraints described above.

Many studies have previously considered so-called hidden sector particles and models, e.g. \cite{Feldman:2007wj, Pospelov:2007mp, Feng:2008mu,ArkaniHamed:2008qn,Pospelov:2008jd, Feng:2009mn, Cohen:2010kn}. Most of these are bottom-up studies of models that may be difficult to embed in String/$M$-theory. Here we make an effort to focus on examples of hidden sectors which are likely to be ultraviolet (UV) complete in String/$M$-theory.

The remainder of this paper is organised as follows. In Section~\ref{cosmo} we review dark matter production in non-thermal cosmologies, and categorise viable DM models arising in string/$M$-theory. In Section~\ref{kineticmixing} we discuss the kinetic mixing portal, which mediates visible sector-hidden sector interactions, and discuss the lifetime and decay phenomenology of the MSSM LVSP in Section~\ref{decaypheno}. In Section \ref{Detection.SEC}, we study how some of the categories are being probed in direct detection experiments. In Appendix~\ref{Weakly} we give an explicit example of a weakly-coupled hidden sector model which can arise from compactified $M$-theory, and discuss its phenomenological implications. 
We present our conclusions in Section~\ref{conclusions}.

\section{Cosmology of Stringy Hidden Sectors}\label{cosmo}

\subsection{Overview and Thermal History}

The scenario we envision is one in which the universe moves from a radiation dominated phase, into a phase in which the energy density is dominated by the coherent oscillation of a single scalar field. This would be the modulus with the longest lifetime, or (equivalently) the smallest mass. There are many species of moduli fields in any given string construction, but we will here be thinking of geometrical moduli, which are common to all constructions, and thus generic in string theory. For such fields, whose interactions with matter tend to involve Planck-suppressed operators, it is common to parameterize the decay width as
\begin{equation} \Gamma_{\Phi} = \frac{C}{8\pi}\frac{m_{\Phi}^3}{M_{pl}^2} \, ,
\label{gammaphi}
\end{equation}
where $m_{\Phi}$ is the mass of the cosmologically relevant modulus field $\Phi$ and $M_{pl}$ is the (reduced) Planck mass.

When the temperature in the radiation-dominated epoch is such that the condition $H=m_{\Phi}$ is satisfied, the modulus field begins coherent oscillation. This period of matter domination ends when the moduli decay. For the sake of this simplified discussion, let us assume that the modulus decays immediately when $H=\Gamma_{\Phi}$, populating the visible sector as well as the many hidden sectors. To be even more specific, let us assume there is a single hidden sector. We will forthwith denote all visible sector temperatures with unprimed variables, and hidden sector temperatures with primed variables.

It is common in the literature to define a re-heat temperature $T_{RH}$ in terms of Eq.~(\ref{gammaphi}) as
\begin{equation}
T_{RH} \equiv \left(\frac{45}{4 \pi^3 g_*(T_{RH})}\right)^{1/4} \times \sqrt{\Gamma_{\Phi} M_{pl}}\, 
\label{TRH}.
\end{equation}
Here $g_*(T_{RH})$ is a weighted sum of the degrees of freedom in the Standard Model which are relativistic at the epoch when that sector is in equilibrium at $T=T_{RH}$. This quantity is (crudely speaking) the effective temperature at which the thermal bath in the visible sector is created upon the decay of the modulus. Thus $T_{RH}$ is very much a quantity that is specific to the visible sector.  This makes sense, in that the strongest bound on this quantity comes from the successful predictions from BBN, which is specific to our visible sector (and mostly unaffected by hidden sector dynamics). This particular quantity thus privileges the visible sector over the hidden sector, but since it is used throughout the literature, we will often utilize~(\ref{TRH}) to replace dependence on $m_{\Phi}$ with dependence on $T_{RH}$. Constraints from BBN require $T_{RH} \gtrsim 1\,{\rm MeV}$. More specifically, for $m_{\Phi} \sim \mathcal{O}(10-100 \, \mathrm{TeV})$, which we will see is motivated by the desire to have weak-scale supersymmetry (SUSY), we must demand $T_{RH} \sim 10 - 100\,{\rm MeV}$. Henceforth we will take $T_{RH} = 10\,{\rm MeV}$ as a benchmark value.

When the modulus decays into the hidden and visible sectors, should we expect them to have the same temperatures? We are assuming these sectors are both radiation dominated immediately after modulus decay, and we are assuming that they interact only very weakly, and thus these two fluids cannot equilibrate with one another. So the question of the relative temperatures depends strongly on the precise identity of the modulus and how it couples to the various sectors of the theory. We will address this model dependence a little later in Subsection~\ref{Weakly-decay}. For the time being let us parameterize the issue by introducing the quantity $\eta$, which represents the fraction of the modulus oscillation energy going to dark radiation, such that $(1-\eta)$ is the fraction going to visible radiation.\footnote{One could distinguish temperatures in the two sectors by considering the distribution of entropy between the two sectors, as opposed to energy density. If the assumption that the universe is radiation dominated at this epoch is valid, then this is a relatively simple conversion. The quantitative dependence on such things as dark matter mass or messenger mass will be unaffected by such changes in conventions.}

With the above in mind, we will define the two temperatures $T_D$ and $T_D'$ as the effective temperature of the radiation in the visible and hidden sectors (respectively) at the moment when $H= \Gamma_{\Phi}$ and the energy density in modulus oscillation instantly converts into particle degrees of freedom. Taking into account the distribution of energy density amongst the two sectors, we may relate $T_D$ and $T_D'$ via
\begin{eqnarray}
T_D &\approx & T_{RH} (1-\eta)^{1/4} \\
\label{TD}
T_D' &\approx& \eta^{1/4} \left( \frac{g_\star(T_D)}{g_\star(T_D')}\right) T_{RH}\, .
\label{TDprime}
\end{eqnarray}
Note that we should expect these temperatures to be at most $\mathcal{O}({\rm MeV})$. These quantities will be relevant as we consider the mechanisms by which hidden sector dark matter can form an order unity fraction of the dark matter we observe in the cosmos.

%%%%%%%%%%%%%%%%%%%%%%%%%%%%%%%%%%%%%
\subsection{Categorisation of Dark Matter Production Mechanisms}
\label{DMmech}

In this section, we briefly summarize the results of~\cite{Kane:2015qea}. 
It is possible to group the disparate dark matter production channels into two classes, defined by the annihilation cross-section for the hidden sector dark matter particle, which we will denote with the symbol $X$. We declare the annihilation to be {\em efficient} if the rate of annihilation is greater than the rate at which dark matter particles are injected into the hidden sector heat bath, either through decays of the LVSP into $X$, or from decays of the modulus field $\Phi$. Since the source of the LVSP is also through modulus decay, the two sources ultimately can be tied to the parameters of the modulus field itself. 
This critical annihilation cross-section is found to be
\begin{align}
\nonumber \langle \sigma v \rangle_{\rm crit} &\approx 4\times 10^{-10} \, \mathrm{GeV}^{-2} \times {B_{X}}^{-1} \left(\frac{m_{\Phi}}{50 \, \mathrm{TeV}}\right) \left(\frac{10 \, \mathrm{MeV}}{T_{RH}}\right)^2 \\
&\approx 4.7 \times 10^{-27} \mathrm{cm}^3 \mathrm{s}^{-1} \times {B_{X}}^{-1} \left(\frac{m_{\Phi}}{50 \, \mathrm{TeV}}\right) \left(\frac{10 \, \mathrm{MeV}}{T_{RH}}\right)^2\, , \label{sigmavcrit.EQ}
\end{align}
where $B_X$ is the branching ratio for $\Phi$ decay into the hidden sector dark matter candidate $X$. 

\begin{figure}[t]
\centering
\includegraphics[scale=0.7]{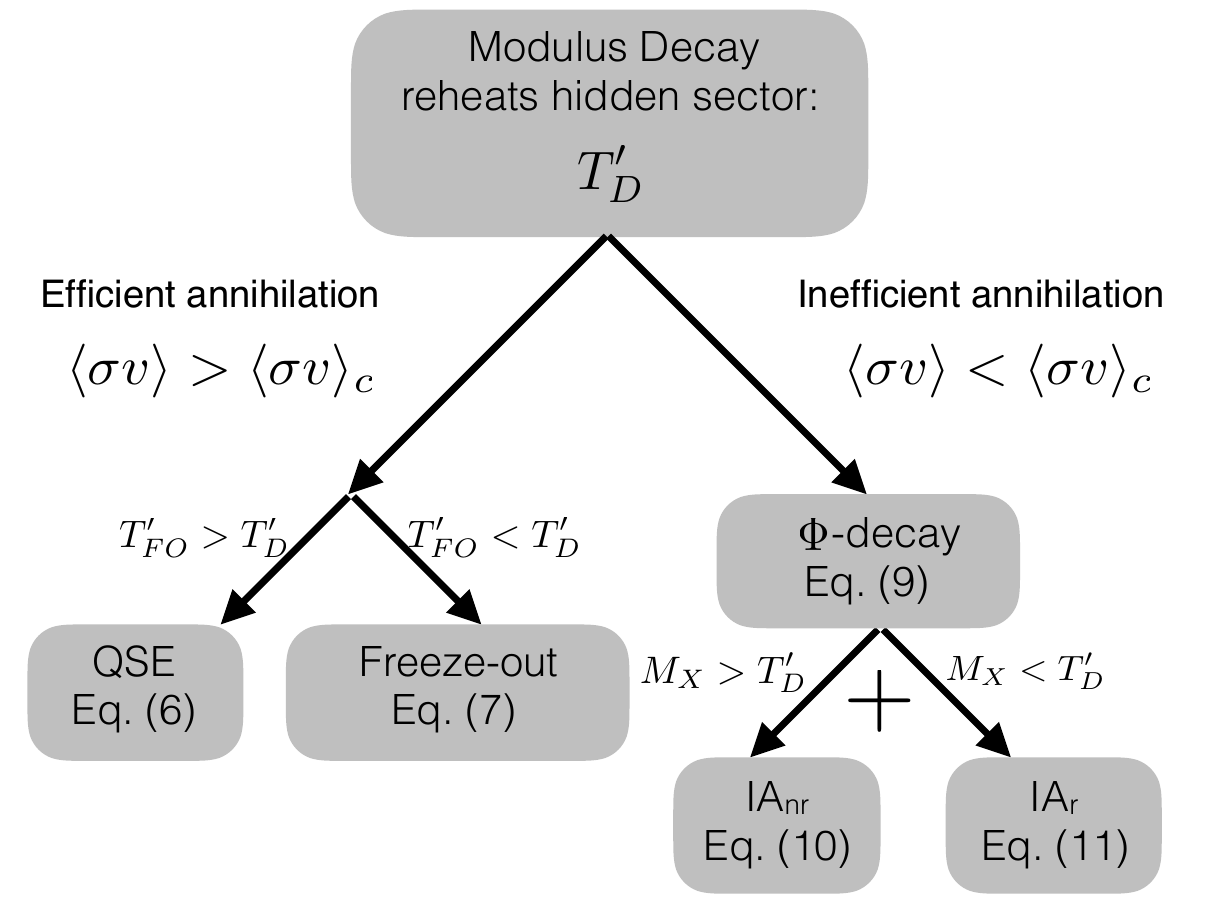}
\caption{Flowchart showing the different categories of dark matter production mechanisms.}
\end{figure}

\subsubsection{Efficient annihilation}

When $\langle \sigma v \rangle|_{T_D'} > \sigv_{\rm crit}|_{T_D'}$ we are in the regime of efficient annihilation. We can further divide this case by examining the dark matter freeze out temperature (as measured relative to the hidden sector heat bath), $T_{\rm FO}'$, defined such that $n^{\rm eq}_{X} (T_{\rm FO}') \sigv \equiv H (T_{\rm FO}')$, where $n^{\rm eq}_X$ is the equilibrium number density of the species $X$.

\begin{itemize}
\item $T_{\rm FO}' > T_D'$: \textbf{Non-relativistic quasi-static equilibrium solution}

If the processes of production and depletion of the dark matter $X$ are both faster than the Hubble expansion rate, then the abundance of $X$ reaches a quasi-static equilibrium (QSE) value, even though the dark matter particles never fully thermalize with the dark radiation. The relic abundance is then given by
\beq
{\Omega h^2}_{\rm QSE} \simeq 0.5  \times (1-\eta)^{-3/4} \left( \frac{M_X}{1 \text{ GeV}} \right)\left( \frac{10 \text{ MeV}}{T_{RH}} \right)\left( \frac{10^{-8} \text{ GeV}^{-2}}{\sigv}\right) \ .
\label{QSE.EQ}
\eeq
Of note is the dependence on the modulus energy fraction going to the visible sector, and the apparent lack of strong dependence on the branching ratio $B_X$ of the modulus decay to $X$, due to the efficiency of annihilation\footnote{There is mild dependence on $B_X$ hidden in the numerical pre-factor. For the full expression giving the exact dependence on $B_X$, see Eq. (37) of \cite{Kane:2015qea}}. This is the generalised version of the non-thermal WIMP miracle \cite{Moroi:1999zb, Acharya:2009zt}.

\item $T_{\rm FO}' < T_D'$: \textbf{Freeze-out during radiation domination}
 
In this case the dark matter particles achieve thermal equilibrium with the heat bath in the hidden sector, freezing out at a later time. This is the standard thermal freeze-out mechanism. The resulting relic abundance depends, as usual, on whether the dark matter particle $X$ is relativistic at the time of freeze-out or not. Defining the ratio $x_{\rm FO}\equiv M_X/T'_{\rm FO}$, we distinguish the two extremes by the usual rule of thumb that $x_{\rm FO} = 3$ divides the two cases:
 \beq
 {\Omega h^2}_{\rm FO^{\rm rad}} \simeq \begin{cases} 0.13 \times \left( \frac{\eta}{(1-\eta)^3 g_\star(T_{\rm FO}) g_\star'(T_{\rm FO}')}\right)^{1/4} \left( \frac{x_{\rm FO}}{17.5}\right)\left( \frac{10^{-8} \text{ GeV}^{-2}}{\sigv}\right) \  \text{if } \ x_{\rm FO} \gtrsim 3, \\ 100 \times \left( \frac{\eta^3}{(1-\eta)^3 g_\star(T_{\rm FO}) g_\star'(T_{\rm FO}')^3}\right)^{1/4}\left( \frac{M_X}{1\,{\rm keV}} \right) \qquad \qquad \qquad \text{if } \ x_{\rm FO} \lesssim 3 \ .
 \label{FOrad.EQ}
 \end{cases}
 \eeq
We note that the condition $T_{\rm FO}' < T_D'$ implies a freeze-out temperature of order 1~MeV or less, and thus the relativistic case is only accessible to relatively low-mass dark matter candidates. Some candidates in this category could be classified as ``warm dark matter'', and therefore suffer/benefit from the same ailments/virtues. 
\end{itemize}

\subsubsection{Inefficient annihilation}

The second class occurs when $\langle \sigma v \rangle|_{T_D'} < \sigv_{\rm crit}|_{T_D'}$, which we designate as \textit{inefficient annihilation}. Here, the dark matter is populated both directly by modulus decay, and by either a) freeze-out during modulus domination (FO$^{\rm mod}$), or by b) dark radiation $\to$ dark matter ``inverse annihilation" (IA). The relic density is then given by the sum of these pieces
\beq
\Omega h^2 = {\Omega h^2}_{\rm \Phi-decay} + {\Omega h^2}_{{\rm FO}^{\rm mod}/{\rm IA}} \ .
\eeq
The relic density contribution from modulus decay is given by
\beq
{\Omega h^2}_{\rm \Phi-decay} \simeq 0.31 \times \left(\frac{B_{X}}{(1-\eta)^{3/4}}\right) \left(\frac{M_{X}}{10 \, \mathrm{MeV}} \right) \left(\frac{T_{RH}}{10 \, \mathrm{MeV}} \right) \left( \frac{ 50 \, \mathrm{TeV}}{m_{\Phi}}\right) \ ,
\label{modDec.EQ}
\eeq
and is common for cases a) and b) above.

The parameter space in which freeze-out during modulus domination occurs is quite restricted (see for example Fig. 6 in \cite{Kane:2015qea}), so we will not discuss it in detail here, as none of the models we will consider fall in that category. 

For inverse annihilation of dark radiation to dark matter, there are two cases:

\begin{itemize}
\item $M_X > T_D'$: \textbf{Non-relativistic case}

The inverse annihilation occurs when $X$ is non-relativistic, and the relic density is given by 
\beq
{\Omega h^2}_{\rm IA_{\rm nr}} \simeq 0.62 \left( \frac{\eta^3}{(1-\eta)^{3/4}g_\star'(T'_\star)^3}\right)\left( \frac{T_{RH}}{10 \text{ MeV}}\right)^7 \left( \frac{1 \text{ GeV}}{M_X} \right)^5\left( \frac{\sigv}{10^{-16} \text{ GeV}^{-2}}\right) \ , 
\label{IANR.EQ}
\eeq
where $T'_\star \sim 0.28 M_X$, which is the point at which the inverse annihilation contribution to the relic density is peaked. This expression holds unless the mass of the dark matter candidate $M_X$ is very large, in which case it becomes exponentially suppressed by a factor $\sim\exp (-2M_X/T')$.

\item $M_X < T_D'$: \textbf{Relativistic case}

The inverse annihilation occurs when $X$ is relativistic, and the relic density is given by 
\beq
{\Omega h^2}_{\rm IA_{\rm r}} \simeq 0.095 \left( \frac{\eta^{3/2}}{(1-\eta)^{3/4}g_\star'(T'_D)^{3/2}}\right) \left( \frac{T_{RH}}{10 \text{ MeV}}\right) \left( \frac{M_X}{1\, \text{keV}} \right) \left( \frac{\sigv}{10^{-19} \text{ GeV}^{-2}}\right) \ ,
\label{IAR.EQ}
\eeq
which clearly requires a very light dark matter candidate, or a very small annihilation rate $\sigv$ in order to give the correct relic abundance.
\end{itemize}

%%%%%%%%%%%%%%%%%%%%%%%%%%%%%%%%%%
\subsection{Categorization of Hidden Sectors}\label{topdown}

Given the discussion in the preceding section, we can begin to categorise cosmologically interesting, weakly-coupled, hidden sectors by invoking general top-down arguments. For concreteness, we will focus on supersymmetric theories where SUSY breaking is gravitiationally mediated to the visible sector. In such theories, the mass scales which can naturally arise in hidden sector models are rather limited. If the hidden sector contains no confining gauge groups, one expects the following:
\begin{itemize}
\item Scalars: gravity mediated SUSY breaking gives all scalar superpartners masses of $\mathcal{O}(m_{3/2})$, where $m_{3/2}$ is the mass of the gravitino. General arguments in supergravity suggest that we should expect $m_{\Phi} \simeq m_{3/2}$, for at least one modulus field, unless a tuning occurs or some form of sequestration can be engineered\cite{Denef:2004cf, GomezReino:2006dk, GomezReino:2006wv, Acharya:2010af}. Given the constraints arising from BBN, this would suggest that scalars in all sectors should have masses of order 50~TeV.\footnote{Throughout this section we are assuming that supersymmetry breaking is communicated to both the visible sector and the hidden sectors, including that in which the dark matter resides, through gravitational-strength interactions. This is generic for string/$M$-theory models in which supersymmetry breaking is associated with a modulus field.}
\item Gauge bosons:  will either be massless, or receive $\mathcal{O}(g_H v_H)$ masses due to symmetry breaking, where $g_H$ and $v_H$ are the gauge coupling and vacuum expectation value in the hidden sector. Given our declaration that only weakly-coupled hidden sectors are to be considered in this work, it is natural to assume that gauge couplings are of order those in the visible sector. Thus we may assume that the ``weak-scale'' in the hidden sector is not dramatically different from that in the visible sector.
\item Gauginos: gaugino masses depend more strongly on the form of modulus stabilization. If moduli appearing in the relevant gauge kinetic function also participate directly in SUSY breaking, one expects gaugino masses of $\mathcal{O}(m_{3/2})$. In all other cases, including those of most phenomenologically promising string constructions, gaugino masses recieve loop-suppressed contributions from both moduli stabilization and anomaly mediation, resulting in $M_{\,\mathrm{gaugino}} \sim m_{3/2}/16\pi^2$~\cite{Dine:1981gu, Randall:1998uk, Giudice:1998xp,Gaillard:1999yb,Acharya:2008zi, Kaufman:2013pya, Kaufman:2013oaa, Everett:2015dqa}.  

\item Chiral fermions: masses are determined by Yukawa couplings, given by $\mathcal{O} (Y_X v_H)$, and can be considerably lighter than $m_{3/2}$.
\end{itemize}

\begin{figure}[t]
\centering
\subfloat[]{
\includegraphics[scale=0.6,valign=c]{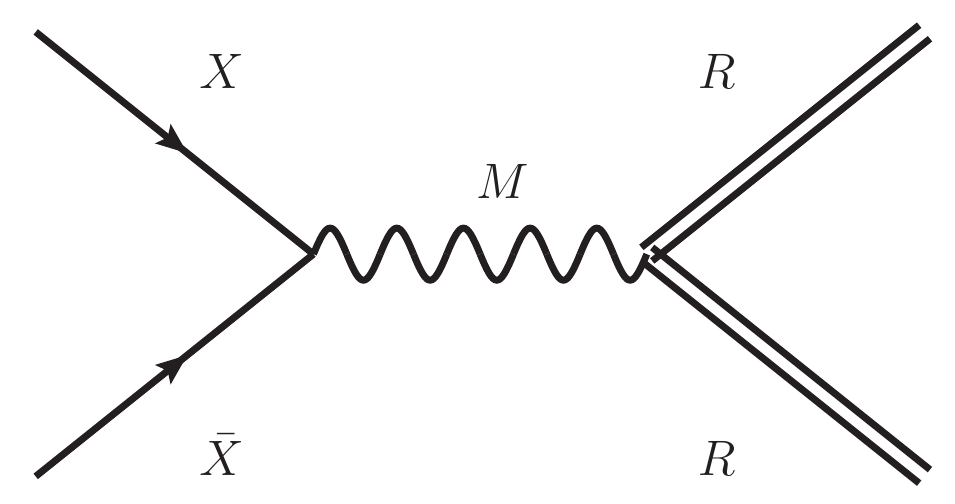}}
\hspace{2 cm}
\subfloat[]{\includegraphics[scale=0.6,valign=c]{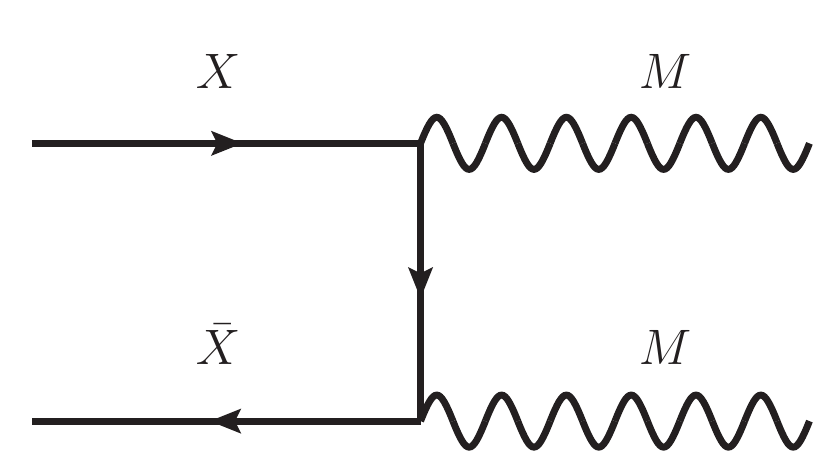}}
\caption{\textbf{a)} Annihilation of $X$ to dark radiation (represented by double lines) when $M_X < M_M$. This occurs in most of the parameter space we consider. \textbf{b)} Annihilation of $X$ directly to the mediator when $M_X \gg M_M$.}
\label{Annihilation.FIG}
\end{figure}

It is now straightforward to argue that for generic hidden sectors with no confining gauge groups, only fermions are likely to give an adequate explanation of dark matter, for natural choices of parameters. 
To see this, note that $\left<\sigma v \right>$, for a particle $X$ annihilating to dark radiation through some mediator $M$, parametrically takes the form
\begin{align}
 \nonumber \sigv \sim \frac{g_D^4}{8 \pi} \frac{M_{X}^2}{M_M^4}\, , M_X < M_M \ , \\
 \sigv \sim \frac{g_D^4}{8 \pi} \frac{1}{M_X^2}\sqrt{1-\frac{M_M^2}{M_X^2}}\, , M_X \gg M_M \ ,
 \label{sigvXmed}
\end{align}
for annihilation via a gauge boson. We expect the dark radiation to be primarily composed of light hidden sector states that do not make up a significant fraction of the dark matter. Such states are found in the example sector we construct in Appendix \ref{SU5.APP}, and are generally expected to be present. Representative diagrams for these two cases are shown in Fig. \ref{Annihilation.FIG}. 

\begin{figure}[H]
\centering
\subfloat[$B_X=0.1$]{\includegraphics[scale=0.26]{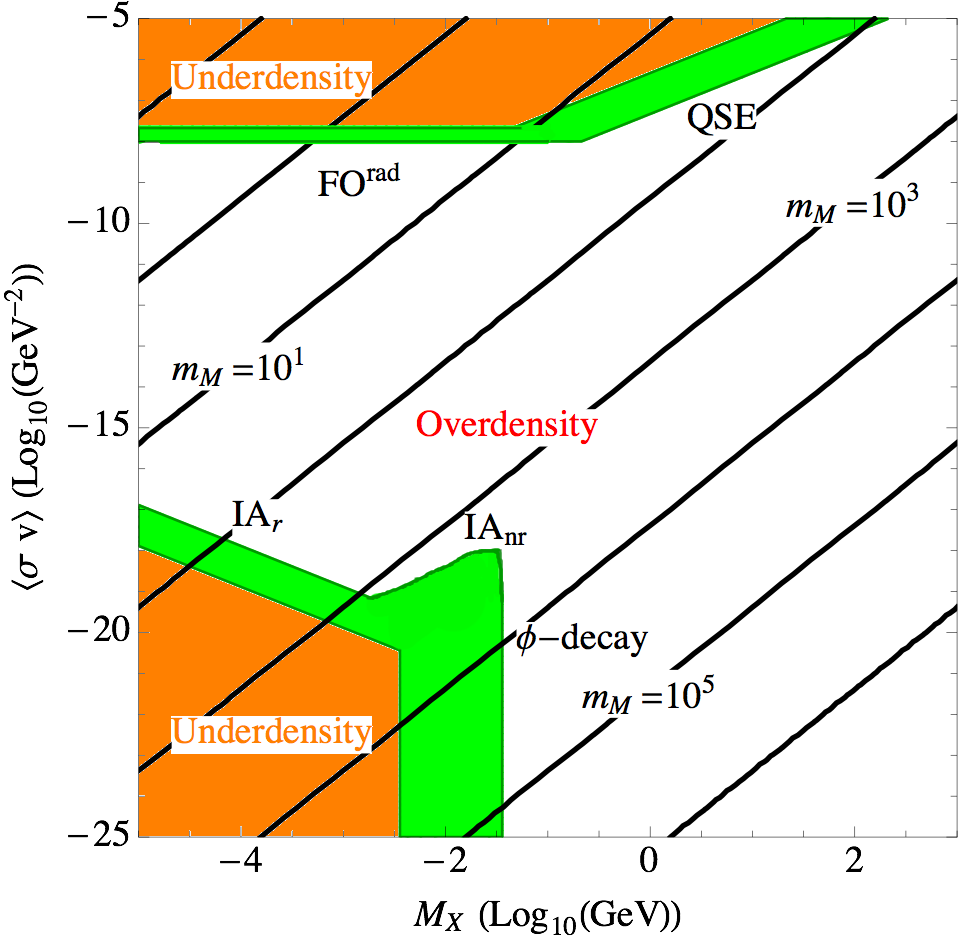} \label{ClassifPlotA.FIG}}
~~
\subfloat[$B_X=10^{-3}$]{\includegraphics[scale=0.26]{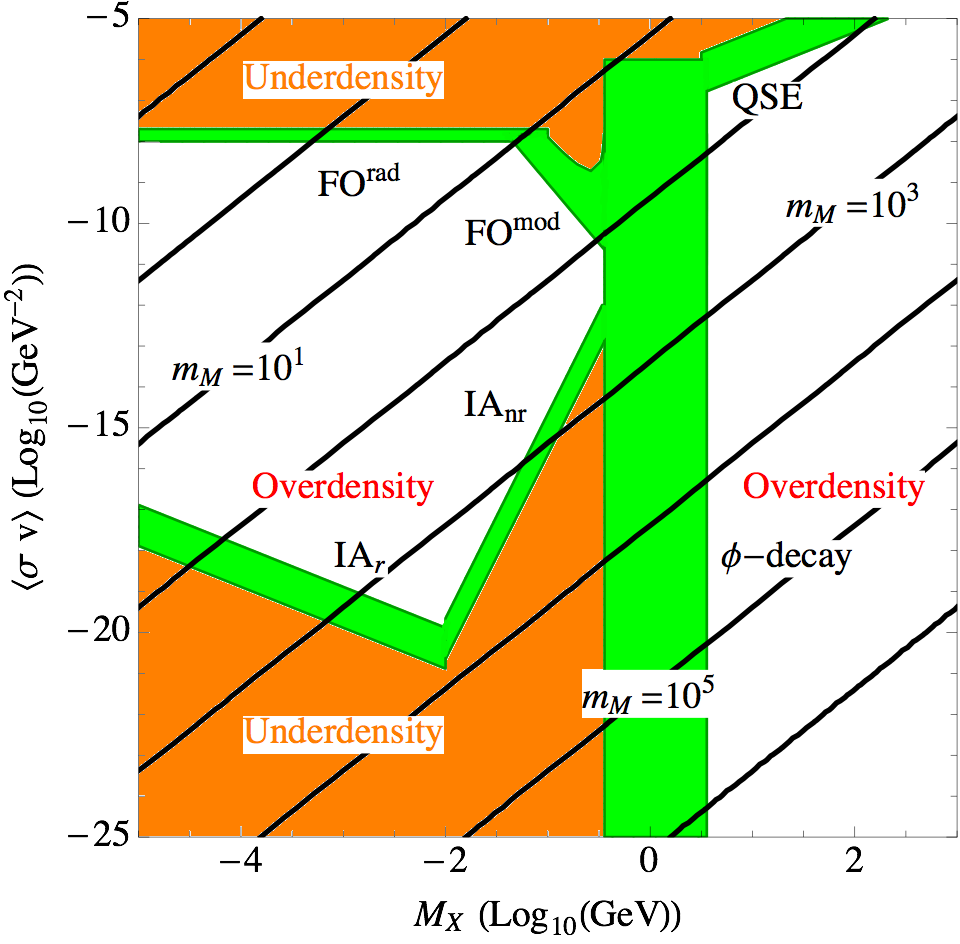} \label{ClassifPlotB.FIG}}
\\
\subfloat[$B_X=10^{-5}$]{\includegraphics[scale=0.26]{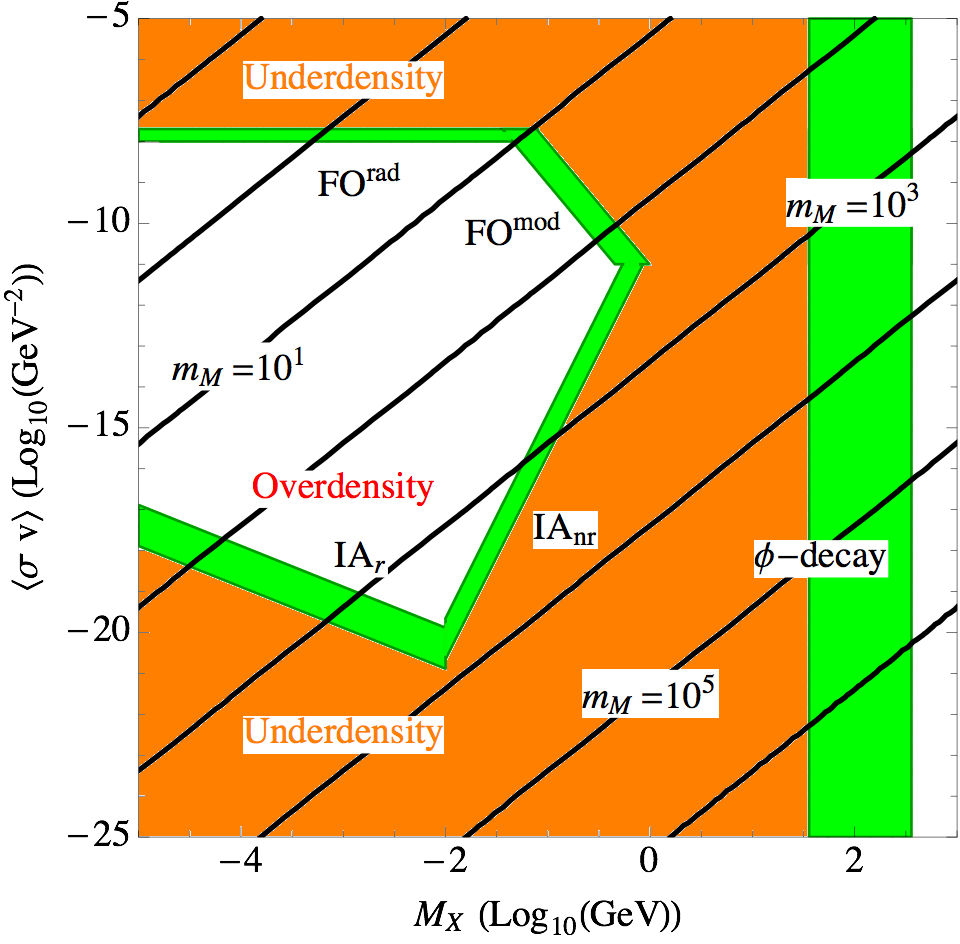}
\label{ClassifPlotC.FIG}}
\caption{The three plots show in green the regions that give $0.012< \Omega_{X} h^2 \leq 0.12$, while in orange $\Omega_{X} h^2 <0.012$. Each plot represents a different choice of the branching ratio $B_X$. In black are contours of constant mediator mass $M_M$ in units of $g_H\times$GeV, calculated using the approximate expressions in Eq. (\ref{sigvXmed}) (only in a very small region in the very top-right corner, $M_M \slashed\propto g_H$). All other parameters are set to the values in Eq. (\ref{BM.EQ}).}
\label{ClassifPlot.FIG}
\end{figure}

By choosing to restrict our attention to weakly-coupled hidden sectors, we can utilize the estimates in~(\ref{sigvXmed}) to replace the pair of quantities $\left(M_X, \langle \sigma v \rangle\right)$ with $\left(M_X, M_M, g_D\right)$, thereby allowing us to identify which sets of values are likely to satisfy the conditions to obtain the correct dark matter abundance for the mechanisms identified in Subsection~\ref{DMmech}.
To make concrete statements, it will be necessary to choose certain benchmark values for the other free parameters in our expressions. For maximal simplicity we will assume the following values
\beq
\eta=0.1,~g_\star=10.75,~g'_\star=10.75,~T_{RH}=10~\text{MeV},~m_\Phi = 50~\text{TeV}\, ,
\label{BM.EQ}
\eeq
and we furthermore assume that the quantities $g_\star,~g'_\star$ are constant at these values throughout the duration of the relevant physics processes. While different choices for these parameters will affect the numerical results of Fig. \ref{ClassifPlot.FIG}, the effects are not substantial. Reducing $g'_\star$ shifts the regions with the correct relic density from (non-)relativistic inverse annihilation to slightly smaller $\sigv$ due to the ${g'_\star}^{-3/2\ (-3)}$ dependence, and has little effect on the regions where freeze-out gives the correct relic abundance (sensitive to ${g'_\star}^{-3/4\ (-1/4)}$ for (non-)relativistic freeze-out). Changing $g'_\star$ does not affect the region where modulus decay produces the relic density at all. Changing $m_\Phi$ must be accompanied by an appropriate change in $T_{RH}$, since they are related by Eq. (\ref{TRH}). Increasing both values commensurately leads to a small shift to larger allowed values of $M_X$ for the regions where quasi-static equilibrium (Eq. (\ref{QSE.EQ}), inverse annihilation (Eqs. (\ref{IANR.EQ}), (\ref{IAR.EQ})) and modulus decay (Eq. (\ref{modDec.EQ})) give the correct relic density. The value of $\eta$ chosen is close to the upper limit, as we will discuss in section \ref{Weakly-decay}. As we are seeking order-of-magnitude estimates of relative and absolute mass scales, choosing the above benchmark values will be sufficient for our purposes. 

We can now represent graphically the allowed parameter space for all processes in Subsection~\ref{DMmech} for different choices of the remaining parameters $B_X,~M_X,~M_M$. This is presented in Figure~\ref{ClassifPlot.FIG}.

\subsubsection{Conditions for Efficient Annihilation}

As can be seen in Fig.~\ref{ClassifPlot.FIG}, in order to fall in one of the regions where efficient annihilation sets the relic abundance, i.e. where $\sigv > \sigv_{\rm crit}$, the dark matter candidate mass must fall in the range $10\,{\rm keV} \lesssim M_X \lesssim 100\,{\rm GeV}$, while the mediator must fall in the range $100\,{\rm MeV} \lesssim M_M \lesssim 100\,{\rm GeV}$, as long as $B_X \gtrsim 10^{-3}$. For $B_X \lesssim 10^{-3}$, several mechanisms begin to contribute and compete, as seen in Fig.~\ref{ClassifPlotC.FIG} for $B_X=10^{-5}$.  

More specifically, considering Eq.~(\ref{QSE.EQ}), the QSE solution requires a dark matter mass of approximately $ \Order(10\,{\rm MeV}) \lesssim M_X \lesssim \Order (100\,{\rm GeV})$, annihilating through a mediator with mass approximately  $\Order(10\,{\rm GeV}) \lesssim M_M \lesssim \Order (100\,{\rm GeV})$\footnote{Since the QSE solution has a small region of validitiy in the $M_X > M_M$ regime, the $g_H$ dependence is not as straightforward to enumerate as for other production mechanisms.}.
Therefore it is unlikely that the efficient annihilation mechanisms for setting the relic density are mediated by a scalar, as $m_{3/2}$ is typically in the multi-TeV range. However, it is possible for the mediator to be a gauge boson, as long as $g_H$ or $v_H$ are small enough. This could correspond, for example, to a gauge boson of similar mass to a visible sector $Z$-boson with mass $\Order (10-100\,{\rm GeV})$, implying the existence of a hidden sector hierarchy problem analogous to that in the visible sector. The dark matter candidate $X$ could be either a light chiral fermion, whose mass would then be set by the Yukawa coupling to the hidden sector Higgs, $y_X$, and $v_H$, or a relatively light hidden sector gaugino with mass $M_X \sim m_{3/2}/16\pi^2$. 

\subsubsection{Conditions for Inefficient Annihilation}

For those mechanisms that fall under the category of inefficient annihilation, where $\sigv < \sigv_{\rm crit}$, the various panels in Figure~\ref{ClassifPlot.FIG} suggest that the inverse annihilation regime is not sensitive to the branching fraction $B_X$, while the region of $\Phi$-decay dominance is dependent {\em solely} on this quantity (and the mass of the dark matter itself). When the overall relic abundance is set by modulus decay, the mass of the dark matter is determined in terms of $B_X$ as
\beq
M_{X} \lesssim 10 \, \mathrm{MeV} \times \left(\frac{0.3}{B_{X}}\right) \left(\frac{10\, \mathrm{MeV}}{T_{RH}}\right)\left(\frac{m_{\Phi}}{50 \, \mathrm{TeV}}\right)  \, .
\eeq
We thereore identify the following three parameter regimes:
\begin{itemize}
\item \textbf{$\mathbf{B_X =0.1}$:}  From Fig.~\ref{ClassifPlotA.FIG}, we see that $M_X \sim 10\, {\rm MeV}$, and the mediator must fall in the range $3\,{\rm TeV} \lesssim M_M/g_H \lesssim 100\,{\rm TeV}$. This could be compatible with a chiral fermion dark matter candidate mediated by either a scalar or a gauge boson. 
\item \textbf{$\mathbf{B_X =10^{-3}}$:} From Fig.~\ref{ClassifPlotB.FIG}, we see that $M_X \sim 1\, {\rm GeV}$, and the mediator mass must fall in the range $50\,{\rm GeV} \lesssim M_M/g_H \lesssim 10^5\,{\rm GeV}$. This is also compatible with a chiral fermion dark matter candidate mediated by either a scalar or a gauge boson. 
\item \textbf{$\mathbf{B_X = 10^{-5}}$:} We see from Fig.~\ref{ClassifPlotC.FIG} that the dark matter candidate can be as heavy as $M_X \sim 100\,{\rm GeV}$, allowing for the possibility that it is a relatively light hidden sector gaugino or a gauge boson. The mediator mass must fall in the range $100\,{\rm GeV} \lesssim M_M/g_H \lesssim 10^5\,{\rm GeV}$, so it could be either a gauge boson or a scalar.
\end{itemize}

For the case where inverse annihilation dominates, we see that a natural division arises for dark matter masses around $10\,{\rm MeV}$, largely independent of $B_X$, for the paratmeter values we have chosen. More specifically:
\begin{itemize}
\item \textbf{IA$_r$:} If the relic abundance is to be set by ``inverse annihilation" of dark radiation to dark matter, $M_X\lesssim 10\,{\rm MeV}$, for relativistic inverse annihilation (Eq. (\ref{IAR.EQ})), and $M_M/g_H \lesssim 7\,{\rm TeV}$. This is compatible with a light chiral fermion dark matter candidate, and a gauge boson mediator. 
\item \textbf{IA$_{nr}$:} For small $B_X \lesssim 10^{-3}$, non-relativistic inverse annihilation gives the correct relic density for  $10\,{\rm MeV} \lesssim M_X \lesssim 500\,{\rm MeV}$ and $ 100\,{\rm GeV} \lesssim M_M/g_H \lesssim 7\,{\rm TeV}$. The heavier mediator would correspond to a lighter dark matter candidate. This would also be compatible with a light chiral fermion dark matter candidate and a gauge boson mediator.
\end{itemize}

An important result is that the dark matter candidate cannot be a scalar, with mass set by $m_{3/2}$, in any region of the parameter space considered. Only for $B_X =10^{-5}$ was it possible that a gauge boson with SM-like values of $g_H$ and $v_H$ could be the dark matter. Additionally, only in two cases is it possible that a light hidden sector gaugino could be the dark matter, namely when annihilation is efficient through the QSE solution, or if $B_X$ is very small,  $B_X\simeq10^{-5}$. The regions where a gauge boson or a light gaugino could be dark matter are small compared to the full parameter space. This strongly suggests that if dark matter is in a hidden sector, \textit{it is likely a fermion whose mass is set by a small Yukawa coupling}. The range of possible outcomes for weakly-coupled hidden sectors is summarized in Table~\ref{Mech.TAB}.

\begin{table}[H]
%\resizebox{\textwidth}{!}{
\centering
\begin{tabular}{c | c | c }
\textbf{Mechanism} & \textbf{Candidate} & \textbf{Mediator}  \\
\hline
&&\\
QSE (Eq. (\ref{QSE.EQ})) & \textit{chiral fermion, (light) gaugino} & \textit{gauge boson} \\
& $0.1 \lesssim M_X \lesssim 100$ GeV & $10 \lesssim M_{Z'} \lesssim 100$ GeV \\
FO$^{rad}$ (Eq. (\ref{FOrad.EQ})) & \textit{chiral fermion}  & \textit{gauge boson}  \\
& $M_X \lesssim 100$ MeV & $M_{Z'} \lesssim 10$ GeV  \\
&&\\
\hline
&&\\
Modulus decay (Eq. (\ref{modDec.EQ})) & \textit{chiral fermion} & \textit{gauge boson, scalar} \\
$B_X\sim 10^{-1}$ & $100 \lesssim M_X \lesssim 500$ MeV  & $1 \lesssim M_{M} \lesssim 100$ TeV \\
 & \textit{chiral fermion} & \textit{gauge boson, scalar}   \\
$B_X\sim 10^{-3}$ & $0.5 \lesssim M_X \lesssim 5$ GeV  & $10 \lesssim M_{M} \lesssim 10^6$ GeV \\
& \textit{chiral fermion, gaugino} & \textit{gauge boson, scalar} \\
$B_X\sim 10^{-5}$ & $50 \lesssim M_X \lesssim 500$ GeV  & $10 \lesssim M_{M} \lesssim 10^6$ GeV  \\
&&\\
\hline
&&\\
IA$_{nr}$ (Eq. (\ref{IANR.EQ})) & \textit{chiral fermion}  & \textit{gauge boson} \\
 & $10 \lesssim M_X \lesssim 100$ MeV & $100 \lesssim M_{Z'} \lesssim 10^4$ GeV \\ 
IA$_r$ (Eq. (\ref{IAR.EQ})) & \textit{chiral fermion}  & \textit{gauge boson} \\
 & $M_X \lesssim 5$ MeV & $10 \lesssim M_{Z'} \lesssim 5000$ GeV \\ 
 &&\\
 \hline
\end{tabular}
%}
\caption{Table summarising the various relic density production mechanisms, with suggested dark matter and messenger candidate masses.}
\label{Mech.TAB}
\end{table}

Note that a conceivable loophole to the above argument arises if $X$ annihilates through a massless gauge boson. We will not consider this possibility further, as DM charged under an unbroken $U(1)$ is strongly constrainted by halo ellipticity constraints~\cite{Feng:2009mn} along with milli-charged DM constraints~\cite{McDermott:2010pa} if the hidden $U(1)$ mixes with $U(1)_Y$.

\subsection{Dependence on Specifics of the String Construction}
\label{Weakly-decay}

In Subsection~\ref{topdown} we were able to suggest possible values for the dark matter mass $M_X$, and mediator mass $M_M$, by utilising~(\ref{sigvXmed}) to determine $\langle \sigma v \rangle$ and thus the relic density $\Omega_{X} h^2$. The expressions in Subsection~\ref{DMmech} were normalized in such a way that our benchmark values of $T_{RH} = 10\,{\rm MeV}$ and $m_{\Phi} = 50\,{\rm TeV}$ will result in the correct value $\Omega_{X} h^2 \simeq 0.1$ for typical models.

However, these expression contain dependence on the precise nature with which the lightest modulus $\Phi$ couples to all sectors in the low-energy effective theory. This occurs through the ratio $\eta$ and the (related) branching fraction $B_X$, as well as through the relative numbers of degrees of freedom in the various sectors, given by the $g_{\star}$ values. Given the types of weakly-coupled models we consider, and the mass scales that are relevant for obtaining $\Omega_{X} h^2 \simeq 0.1$, it is perhaps reasonable to assume that $g'_{\star}$ will be similar to that of the visible sector at temperatures of order 1~MeV or less. The issue of how entropy and energy is distributed between the various sectors is potentially more troublesome.

From Eq.~(\ref{sigmavcrit.EQ}) we see that the $\langle \sigma v \rangle_{\rm crit}$ depends directly on the branching ratio $B_X$ into the hidden sector particle $X$. An accurate determination of $B_X$ requires knowledge of how the modulus decays to populate all sectors, including the visible sector and all hidden sectors. This depends on details such as the number of hidden sectors and the couplings to each sector, all of which depend on the nature of the compactification that is being considered. A rough estimate is that $B_X \sim 1/ (N_{\rm sectors}\times N_{\rm d.o.f.})$, where $N_{\rm d.o.f.}$ is the number of degrees of freedom in each if the various sectors. The number of degrees of freedom in the Standard Model or MSSM is $\mathcal{O}(100)$, and we typically expect many such sectors. Thus, it is reasonable to assume that $B_X$ is significantly less than unity, but it is difficult to be more precise from the top-down perspective.

From the bottom-up, however, we know that the successful predictions of BBN and the formation of the cosmic microwave background (CMB) are both sensitive to the Hubble parameter during those epochs. This, in turn, puts a limit on the total number of relativistic degrees of freedom in the cosmos at those times. The limits are often expressed in terms of limits on the number of effective neutrino species, or sometimes just the excess over the canonical three generations of neutrinos in the Standard Model, $\Delta N_{\rm eff}$. The current bounds are  $\Delta N_{\rm eff} (T_{\rm BBN}) \lesssim 1.44$ \cite{Fields:2006ga} and $\Delta N_{\rm eff}(T_{\rm CMB})\lesssim 0.4$ \cite{Talk:2015}, where $T_{\rm BBN} \approx 1$ MeV and $T_{\rm CMB} \approx $ 1 eV. These bounds can be translated into constraints on the ratio of visible and hidden sector temperatures \cite{Feng:2008mu}:\begin{equation}\label{neff}
g^\prime_*(T^\prime_{\rm BBN}) \left(\frac{T^\prime_{\rm BBN}}{T_{\rm BBN}}\right)^4 \lesssim 2.52, \,\,\, g^\prime_*(T^\prime_{\rm CMB}) \left(\frac{T^\prime_{\rm CMB}}{T_{\rm CMB}}\right)^4 \lesssim 0.18\, ,
\end{equation} where $g^\prime_*(T^\prime)$ is the effective number of relativistic degrees of freedom in the hidden sector.

These limits have been explored in a number of studies of the so-called ``dark radiation'' problem, which can be very constraining on string-derived models \cite{Hebecker:2014gka,Allahverdi:2014ppa,Cicoli:2015bpq, Acharya:2015zfk,Stott:2017hvl}. In the present case, the limits can be interpreted as a bound on the parameter $\eta$, as it effectively sums over the branching ratio of the lightest modulus to dark radiation in all hidden sectors of the compactified theory, and is potentially large as a result. 
Denoting visible sector relativistic particles as $R_i$ and hidden sector relativistic particles as $R^\prime_{i}$, so that $\sum B_{R'_i}=\eta$, we obtain the following relations:
\begin{equation}\label{entropy}
\frac{s^\prime}{s} = \left(\frac{\sum B_{R^\prime_i}}{\sum B_{R_i}}\right) \Rightarrow \frac{\sum B_{R^\prime_i}}{\sum B_{R_i}}  = \frac{g^\prime_*(T^\prime)}{g_*(T)} \left(\frac{T^\prime}{T}\right)^3
\end{equation} 
where $s^\prime$ and $s$ are hidden and visible sector entropy densities. Combining (\ref{neff}) and (\ref{entropy}), we obtain:\begin{equation}
\frac{\sum B_{R^\prime_i}}{\sum B_{R_i}} \lesssim 2.52 \times \frac{g^\prime_*(T^\prime_{\rm BBN})^{1/4}}{g_*(T_{\rm BBN})}, \,\,\, \frac{\sum B_{R^\prime_i}}{\sum B_{R_i}} \lesssim 0.18 \times \frac{g^\prime_*(T^\prime_{\rm CMB})^{1/4}}{g_*(T_{\rm CMB})}
\end{equation} where $g_*(T_{\rm BBN}) \approx 10$ and $g_*(T_{\rm CMB}) \approx 3$. Thus if the hidden sector contains particles with masses $\lesssim$ eV i.e. $g^\prime(T^\prime_{\rm CMB}) \neq 0$, $\Delta N_{\rm eff}$ constraints impose the rather strong bound
$\sum B_{R^\prime_i} \lesssim  0.1 \times \sum B_{R_i}$. This corresponds to the constraints $\eta \lesssim 0.20$ and $\eta \lesssim 0.06$ at BBN and at CMB ,respectively~\cite{Kane:2015qea}. 
Future experiments will have even greater sensitivity to $\Delta N_{\rm eff}(T_{\rm CMB})$, such as CMB-S4 \cite{Abazajian:2016yjj}, and will therefore provide even stronger constraints on hidden sector model building. In our work we will assume these constraints satisfied in the following analysis. Note that our benchmark value of $\eta = 0.1$ is at the upper limit of what is allowed. As has been noted elsewhere, one of the immediate cosmological problems a string-based model must face is the engineering of a preference for populating {\em only} the visible sector at a given epoch -- whether during re-heating at the end of inflation, or upon the decay of the lightest modulus field \cite{Reece:2015lch, Adshead:2016xxj, Tenkanen:2016jic, Hardy:2017wkr}. This is an important problem that certainly deserves more attention.

\section{Kinetic Mixing and MSSM-LSP Decay}\label{decay}

The preceding section described the conditions on the hidden sector(s) such that the relic abundance of any stable particle in such a sector saturate the cosmologically observed dark matter abundance, while remaining consistent with other cosmological constraints. Together, these conditions require that the lightest visible supersymmetric particle (LVSP) -- presumably the MSSM-LSP -- must decay on a cosmologically prompt time-scale. We can parameterize the decay of the LVSP as
\begin{equation}
\Gamma_{\mathrm{LVSP}} \equiv \frac{C_{\mathrm{LVSP}}}{8 \pi} M_{\mathrm{LVSP}} \approx 10 \, \mathrm{s}^{-1} \times \left(\frac{C_{\mathrm{LVSP}}}{6 \times 10^{-23}}\right) \left(\frac{ M_{\mathrm{LVSP}}}{100 \, \mathrm{GeV}}\right) \, ,
\end{equation}
where we have normalized the effective decay coefficient $C_{\mathrm{LVSP}}$ in such a was as to yield a lifetime on the order of $\tau_{\mathrm{LVSP}} \lesssim 0.1$ s, which is just before the epoch of BBN. What sort of operator can generate sufficiently short lifetimes for the LVSP?

Assuming no new mass scales are generated through strong dynamics, then any non-renormalizable couplings would be suppressed by powers of $1/{M_{pl}}$. For an LVSP mass of order 100~GeV, one might estimate a value for $C_{\mathrm{LVSP}}$ as 
\begin{equation} C_{\mathrm{LVSP}} \sim \left(\frac{100\,{\rm GeV}}{M_{pl}}\right)^2 \sim 10^{-34}\, .
\end{equation}
Thus if the visible sector LSP decays before BBN, the portal must proceed through renormalizable dimension-4 operators. There are two primary portals between hidden sectors and the visible sectors considered in the literature, the so-called ``Higgs portal" \cite{Schabinger:2005ei, Patt:2006fw, MarchRussell:2008yu} and the ``kinetic mixing portal" \cite{Holdom:1985ag, Dienes:1996zr}. We will focus here on the latter portal, as this one is theoretically well motivated in our framework, and indeed expected to occur quite generically in string/$M$-theory~\cite{Dienes:1996zr}.

\subsection{The Kinetic Mixing Portal}\label{kineticmixing}

\begin{figure}[th]
\centering
\includegraphics[scale=0.5]{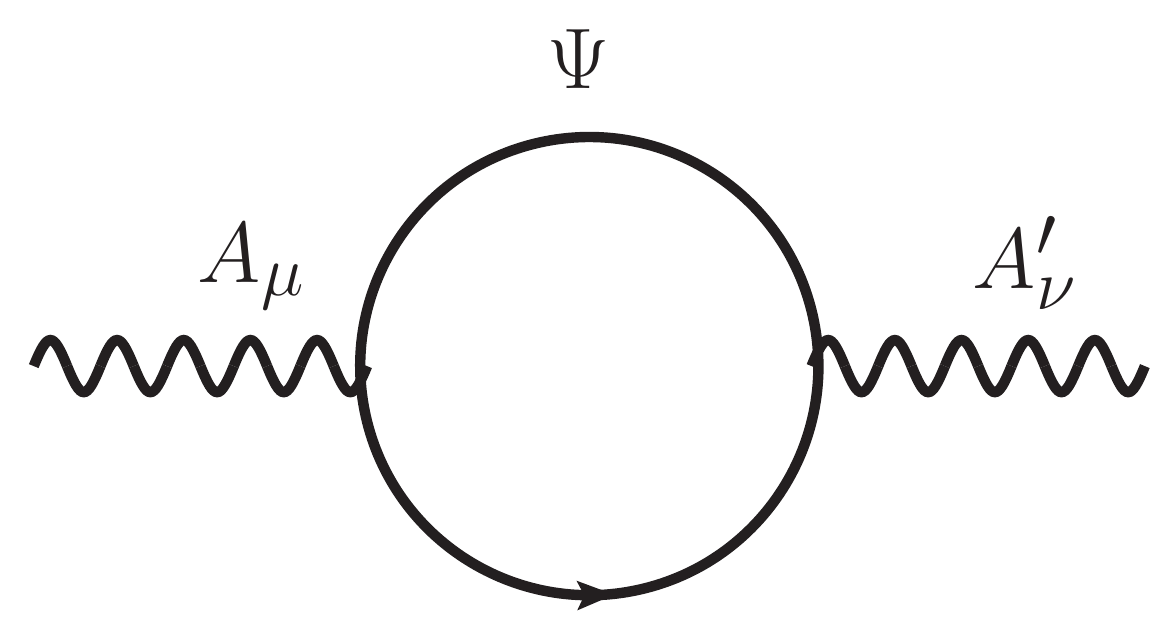}
\caption{Diagram representing the one-loop charge renormalisation that induces a mixing between the visible sector $U(1)$ gauge field $A_\mu$ and the hidden sector $U(1)'$ gauge field $A'_\nu$. The field $\Psi$ is a field charged under both gauge groups.}
\label{KMLoop.FIG}
\end{figure}

The authors recently argued elsewhere \cite{Acharya:2016fge} that any generic string/$M$-theory construction is likely to exhibit kinetic mixing, and thus the lightest visible sector particle (such as the MSSM-LSP) is very unlikely to be stable. 
Kinetic mixing in four-dimensional field theory arises at one-loop order via a diagram involving the visible sector $U(1)$ gauge field, the hidden sector $U(1)$ gauge field, and heavy fields charged under both gauge groups, as shown in Fig.~\ref{KMLoop.FIG}. In open string theories, if the $U(1)$'s are supported by $D$-branes which are separated in the extra dimensions, then massive open strings stretching between the $D$-branes give rise to massive bi-charged fields. This would apply for all supersymmetric Type I, Type IIA and Type IIB models, and there are generalizations of this statement in heterotic string theory, $M$- and $F$-theory.

Writing the gauge part of the Lagrangian as
\beq
\Lagr_{gauge} = -\frac{1}{4} F^{\mu\nu} F_{\mu \nu} - \frac{1}{4}G^{\mu\nu}G_{\mu\nu} + \frac{\epsilon}{2}F^{\mu\nu}G_{\mu\nu}\, ,
\label{Gauge.EQ}
\eeq
where $F^{\mu\nu} = \partial^{[\mu} A^{\nu]}$ is the field strength of the visible sector $U(1)$ and $G^{\mu\nu} = \partial^{[\mu} A'^{\nu]}$ is the field strength of the hidden sector $U(1)$. We see that the one-loop diagram in Fig. \ref{KMLoop.FIG} will give the rise to a non-zero mixing parameter $\epsilon$. The size of $\epsilon$ can be determined by computing the charge renormalisation diagram. The result is
\beq
\epsilon = \frac{g_a g_b}{12\pi^2} \left(Q^a Q^b \right) \left( \log\left(\frac{M^2_{ab}}{M^2}\right)\right)
\eeq
in the case of two $U(1)$ groups $a$ and $b$. The scale $M$ corresponds to the mass of the fields which are charged under both groups, with $M \sim {R \over l_s^2} \sim M_{GUT} \sim 10^{16}$ GeV in most constructions, where $R$ is the separation of two stacks of $D_p$ branes, connected by the open string. The term $M_{ab}$ is the mass matrix of the chiral superfields charged under both groups. 
Since the chiral superfields gain a mass-squared of order $M^2 + \Delta m^2$, where $\Delta m^2$ is the mass-squared splitting due to radiative corrections of the various fields, we may rewrite the expression for $\epsilon$ as
\begin{equation}
\epsilon = \frac{g_a g_b}{6\pi^2} \left(Q^a Q^b \right) \log\left(1+\frac{\Delta m_{ab}}{M}\right)\, .
\label{eps.EQ}
\end{equation}
Thus we see that the size of $\epsilon$ depends very strongly on the ratio $\Delta m_{ab}/M$. If this ratio is $\Delta m_{ab}/M \ll1$, then $\epsilon$ is negligibly small. This is the likely outcome in cases where the breaking from some non-Abelian group to $U(1)$ arises through some vacuum expectation value at low scales, generating splittings of order the gravitino mass ($10^4$ GeV). 

A more interesting alternative, motivated from string/$M$-theory constructions, is the case in which the $U(1)$ arises through Wilson line breaking, which would occur at scales of order the visible sector GUT scale ($10^{16}$ GeV). In this case one expects $\epsilon \sim 10^{-3}$, as opposed to  $\epsilon \sim v/M_{GUT}$ for vacuum expectation value breaking. Requiring the visible LSP to decay before BBN precludes the spontaneous symmetry breaking scenario which generates $\epsilon \sim 10^{-13}$. Thus the only phenomenologically viable option is for the kinetic mixing to be generated via Wilson line breaking. Since a multitude of hidden sectors are expected in string/$M$-theory compactifications, we generically expect at least one hidden sector whose GUT group is broken via a Wilson line.\footnote{This requires that the hidden sector gauge theory be localized on a manifold with a non-trivial first homotopy group.} In this sense, \textit{Wilson line breaking ``picks out" the particular hidden sector into which the visible LSP decays}. We will discuss the implications of this constraint in more detail in the next subsection. 

First, however, we wish to emphasize that we are envisioning a scenario in which the MSSM-LSP decays, yet this is occurring without explicit violation of R-parity (or matter parity) in the visible sector. We note that matter parity (and its generalization to R-parity) is defined in terms of certain charge assignments that are specific to the field content of the MSSM (visible) sector. Such a symmetry, if it is indeed present in the supersymmetric Lagrangian, may be the result of a discrete symmetry with deep string-theoretic origins~\cite{Ibanez:1991pr,Dreiner:2005rd}, or it may be the result of a spontaneously broken gauged $U(1)_{\rm B-L}$, which often arises for example in $SO(10)$ GUT models. In either case, there should be no expectation that a similar discrete symmetry must be operative in the various hidden sectors. In that case, the LVSP can decay to the LHSP and a SM boson. The LHSP can then decay into any fields allowed by the symmetries of the hidden sector, which might well not include an analogue to R/matter-parity.

\subsection{Phenomenological Signatures of a Decaying Visible Sector LSP}
\label{decaypheno}

Since the visible sector LSP is no longer stable, and indeed could have a very short life time, its decay can be searched for at colliders. Signatures of such a decay in a detector have been discussed in~\cite{Arvanitaki:2009hb} in the context of Ramond-Ramond $U(1)$'s mixing with hypercharge $U(1)_{Y}$. 

The gaugino mass Lagrangian in the presence of kinetic mixing of an additional $U(1)'$ and $U(1)_Y$ is
\begin{align}
\nonumber \mathcal{L}_{G.\,mass} &= -\frac{1}{2} M_2 \tilde{W}^a \tilde{W}^a - \frac{1}{2}M_Y \tilde{Y}\tilde{Y}  - \frac{1}{2}M_X \tilde{X}\tilde{X} - M_{XY} \tilde{X}\tilde{Y} + h.c. \, \\
&= -\frac{1}{2} M_2 \tilde{W}^a \tilde{W}^a - \frac{1}{2}M_1 \tilde{B}\tilde{B}  - \frac{1}{2}M_{\tilde{Z}'} \tilde{Z}'\tilde{Z}' - M_{BZ'} \tilde{B}\tilde{Z}' + h.c. \, ,
\end{align} 
where $\tilde{W}^a$ and $\tilde{Y}$ are the $SU(2)_L$ and $U(1)_Y$ gaugino fields of the visible sector, and $\tilde{X}$ is the gaugino field of the hidden sector $U(1)'$. The second line is obtained after diagonalisation of the gauge fields to the canonical basis, and all the effects of the kinetic mixing are encoded in the $U(1)_X$ sector. The diagonalisation is obtained by the rotation~\cite{Choi:2006fz}
\begin{align}
\bmat \tilde{Y} \\ \tilde{X} \emat = \bmat 1 & \epsilon/\sqrt{1-\epsilon^2} \\ 0 & 1/\sqrt{1-\epsilon^2} \emat \bmat \tilde{B} \\ \tilde{Z}' \emat \, ,
\end{align}
where $\epsilon$ is the same kinetic mixing parameter from Eq. (\ref{Gauge.EQ}), such that the mass parameters in the above Lagrangian are given by
\begin{align}
M_1 = M_Y;~~M_{\tilde{Z}'} = \frac{1}{1-\epsilon^2}\left( M_X + \epsilon M_{XY}  + \epsilon^2 M_Y \right);~~ M_{BZ'} = \frac{1}{1-\epsilon^2} \left( M_{XY}  + \epsilon M_Y \right) \ .
\label{MassDef.EQ}
\end{align}
The mixing mass term $M_{XY}$ can be zero at tree level, although it will be generated by radiative effects. As such, we can take it to be small compared with the other bilinear mass terms $M_Y,~M_2$ and $\mu$.
Note that the diagonalisation above also rotates the gauge field strengths so that they are canonically normalised. This procedure does \textit{not} induce a small mass for the Standard Model photon. An intuitive means of understanding that this is the case is that the electroweak symmetry group in the visible sector is broken in the usual way by the Higgs mechanism after the diagonalisation to the canonical basis. The symmetry breaking still leaves behind an unbroken $U(1)_{em}$, which has an associated massless boson, namely the photon. A detailed analysis of the mass eigenstates of the gauge boson sector of such a kinetically mixed theory was performed in~\cite{Babu:1997st, Feldman:2007wj}. Because the hidden sector $U(1)$ is broken to give a mass to the $Z'$, and there is no St\"uckelberg mass mixing of the two $U(1)'s$, the hidden sector fields are not millicharged under $U(1)_{em}$ \cite{Feldman:2007wj, Feldman:2010wy}.

The neutralino mass matrix is now a $5 \times 5$ matrix, given by
\begin{align}
{\cal{M}}_{N} = \left( \begin{array}{c c c c | c}  & & & & M_{BZ'} \\ & & {\cal{M}}_{N,\,4\times4} & & 0 \\ & & & & g_X v c_\beta Q_1 \\  & & & & g_X v c_\beta Q_2 \\ \hline M_{BZ'} & 0 & g_X v c_\beta Q_1 & g_X v s_\beta Q_2 & M_{\tilde{Z}'} \end{array} \right) \, ,
\label{ExtMat.EQ}
\end{align}
where ${\cal{M}}_{N,\,4\times4}$ is the usual MSSM visible sector neutralino mass matrix. The gauge coupling of the $U(1)_X$ sector is $g_X$, $v$ is the usual SM vacuum expectation value, and $\beta$ is the usual MSSM Higgs mixing angle. Since the $U(1)_X$ arises in a true hidden sector, the MSSM Higgs fields $H_1,~H_2$, carry no charge under it at tree level. However, non-zero charges $Q_1$ and $Q_2$ are induced due to the kinetic mixing of $U(1)_Y$ and $U(1)_X$, with
\begin{align}
Q_i = \frac{g_Y}{g_X} \frac{\epsilon}{\sqrt{1-\epsilon^2}} Y_i \, ,
\label{HiggsinoMixing.EQ}
\end{align} 
where $i=1,2$, $g_Y$ is the usual $U(1)_Y$ gauge coupling, and $Y_i$ is the corresponding hypercharge assignment, $\pm1/2$. 

In the MSSM without considering hidden sectors that kinetically mix, if the LSP is a neutralino, it is the lightest eigenstate of the $4\times4$ sub-matrix of Eq.~(\ref{ExtMat.EQ}) above. When the mixing between neutralinos is small, i.e. the mass parameters $M_1, M_2, \mu \gg v$, the lightest neutralino is almost pure Bino/Wino/Higgsino, depending on which mass parameter ($M_1/M_2/\mu$) is smallest. 

When the hidden sector is included as it should, the LSP will now be the lightest eigenstate of the full $5\times5$ matrix in Eq.~(\ref{ExtMat.EQ}). Now in the small mixing case, i.e. $M_1, M_2, \mu,M_{\tilde{Z}'} \gg v$, the LSP will again be an almost pure state, either Bino/Wino/Higgsino, but importantly, also potentially $Z'$-ino. Therefore, if the condition $M_{\tilde{Z}'} < M_i, \mu$ is satisfied, the state which is approximately pure hidden sector $\tilde{Z}'$ will be the lightest neutralino. This would mean the \textit{true} LSP is not a visible sector state, but rather the hidden sector gaugino!

\subsubsection{Decay time of the visible LSP to hidden LSP}

If the mass difference between the visible LSP $\chi_i$ and the hidden LSP $\chi_j$, $M_{\chi_i} - M_{\chi_j} \equiv \delta m$, is greater than the $Z$-boson mass $m_Z$, the visble LSP undergoes 2-body decay to a $Z$-boson with lifetime:
% --- TWO BODY LIFETIME
\begin{align}
\tau^{\chi_i \rightarrow Z \chi_j}_{\rm 2-body} &\sim 10^{-17} ~ \text{s} \times \left( \frac{10^{-3}}{\epsilon} \right)^2 \left( \frac{0.01}{|N_{i3}N_{j3}^*-N_{i4}N_{j4}^*|}\right)^{2} \ ,
\end{align}
where $N_{km}$ is an element of the matrix which diagonalises the enlarged neutralino mass matrix in Eq.~(\ref{ExtMat.EQ}). 
Dependence on the kinetic mixing parameter $\epsilon$ arises due to the dependence on $\epsilon$ of the terms defined in Eqs.~(\ref{MassDef.EQ}, \ref{HiggsinoMixing.EQ}) that are seen in the extended mass matrix. We have chosen $m_{\rm LVSP}=1$ TeV and $m_{\rm LHSP} =100$ GeV as benchmark values. 

\begin{figure}[t]
\centering
\subfloat[]{
\includegraphics[scale=0.6,valign=c]{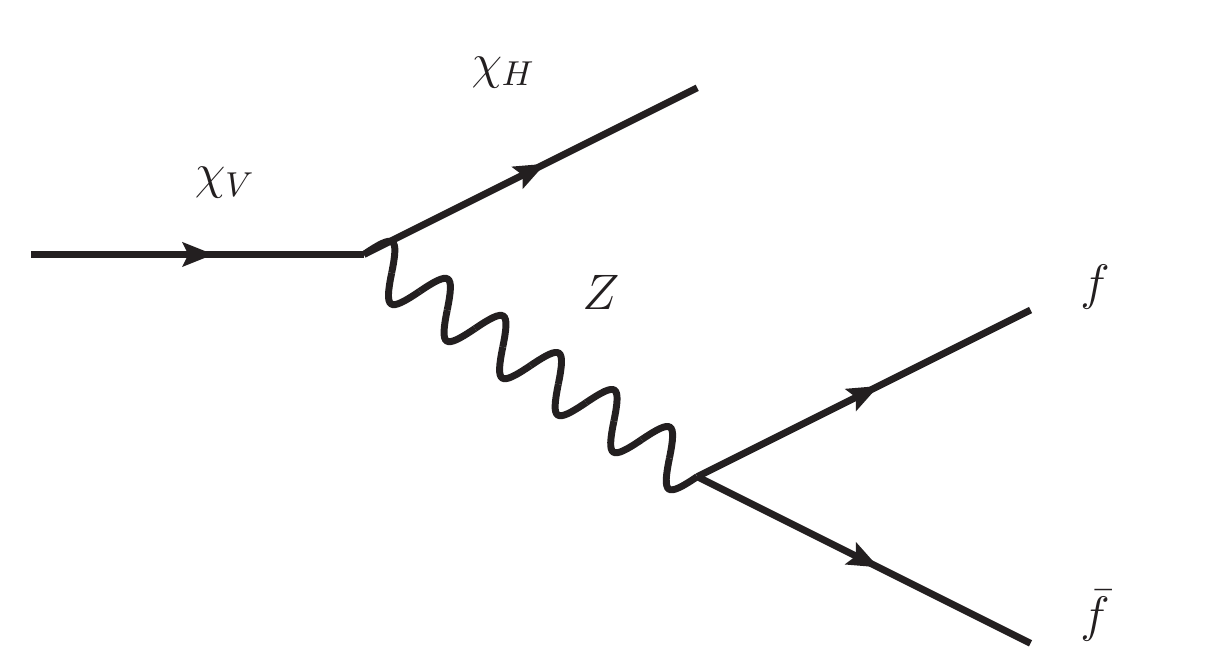}}
\subfloat[]{\includegraphics[scale=0.6,valign=c]{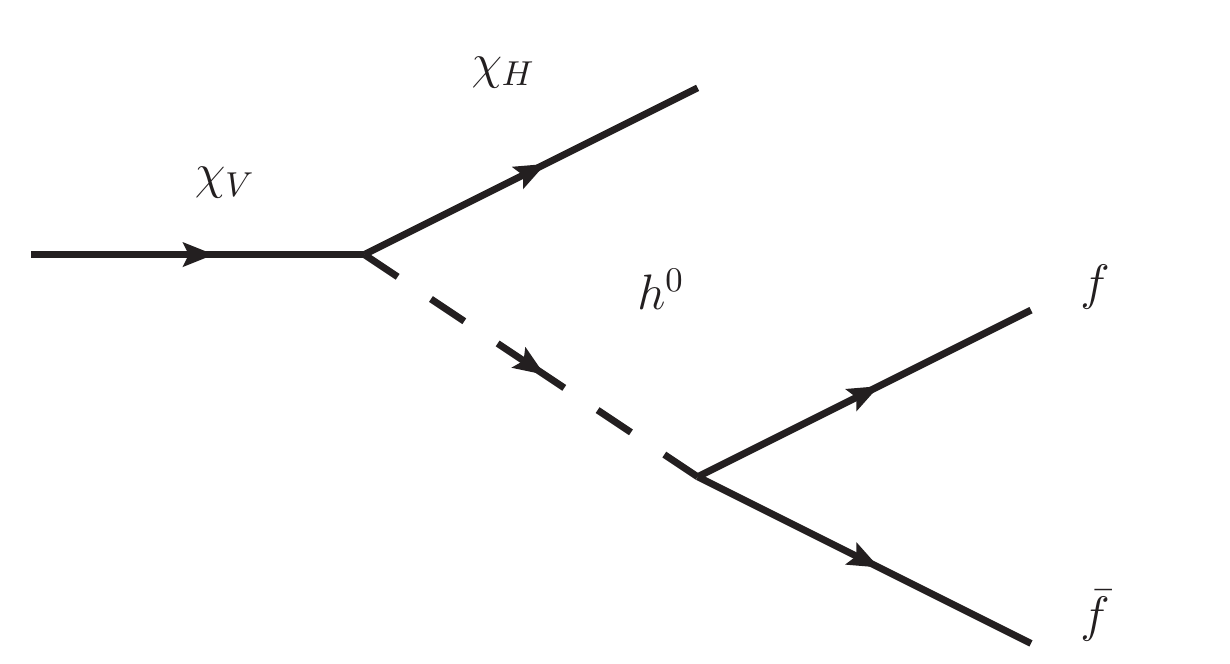}}
\caption{\textbf{a)} Three-body decay of visible LSP $\chi_V$ into hidden gaugino $\chi_H$ and Standard Model fermions via an off-shell $Z$ boson. \textbf{b)} As in a), but via an off-shell Higgs boson.}
\label{LSPdecay.FIG}
\end{figure}

Three-body decays dominate if the mass difference of the two neutralinos is sufficiently small, namely $\delta m < m_Z$~\cite{Arvanitaki:2009hb}. These decays are depicted in Fig. \ref{LSPdecay.FIG}. Indeed, if SUSY breaking is mediated to all sectors in the same way, one might expect that all gaugino masses should be similar in size, such that a small mass splitting is not unlikely. 
The characteristic lifetime in such a scenario can be calculated from~\cite{Djouadi:2001fa} for decays via a $Z$-boson or a Higgs, and is found to be:
\begin{align}
\tau^{\chi_i \rightarrow Z \chi_j}_{\rm 3-body} &\sim 10^{-12}~ \mathrm{s}  \times \left( \frac{10^{-3}}{\epsilon} \right)^2 \left( \frac{0.01}{|N_{i3}N_{j3}^*-N_{i4}N_{j4}^*|}\right)^{2} \left(\frac{\delta m}{50 \ \mathrm{ GeV}}\right)^5 \ ,
\end{align}
for decays via an off-shell $Z$ boson, where we have chosen $m_{\rm LVSP}=1$ TeV and $m_{\rm LHSP} =950$ GeV as benchmark values (shown in Fig. \ref{LSPdecay.FIG}a). Again, $\epsilon$ dependence enters via the diagonalisation of the extended neutralino mass matrix.

The LVSP can also decay to a Higgs boson, again either via a 2-body or a 3-body process, depending on the mass difference $\delta m$, although this mode is sub-dominant. The lifetimes can be obtained by scaling the appropriate expressions
\beq
\tau^{\chi_i \rightarrow h\, \chi_j} \sim \frac{c_w^2 H_{ij}^2}{|N_{i3}N_{j3}^*-N_{i4}N_{j4}^*|^2} \tau^{\chi_i \rightarrow Z\ \chi_j} \ ,
\label{HiggsDecay.EQ}
\eeq 
where $c_w=\cos\theta_w$, and $H_{ij}$ is the neutralino coupling to Higgs bosons, given by
\beq
H_{ij}= \frac{1}{2s_w}\left( N_{j2}-t_w N_{j1}\right)\left(-s_\alpha N_{i3} -c_\alpha N_{i4}\right) + i\leftrightarrow j \ .
\eeq
This decay mode is shown in Fig. \ref{LSPdecay.FIG}b.

\begin{figure}[t]
\centering
\includegraphics[scale=0.8]{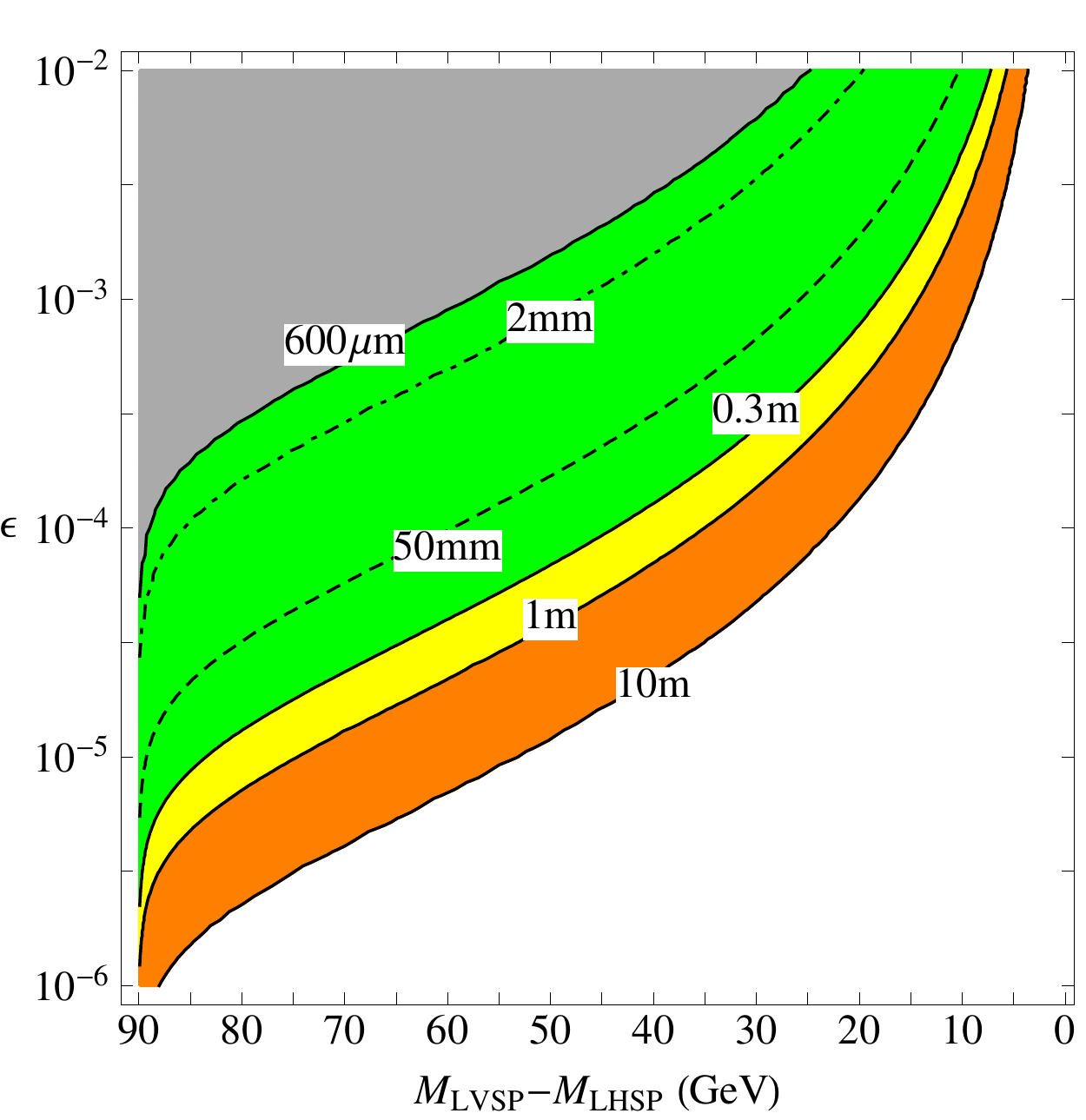}
\caption{Proper decay lengths of the LVSP into an off-shell $Z$-boson and the LHSP for $M_{\rm LVSP} - M_{\rm LHSP} < m_Z$, such that there can be displaced decays. We have chosen $|N_{i3}N_{j3}^*-N_{i4}N_{j4}^*|=0.1$ as a reference value. }
\label{DispV.FIG}
\end{figure}

The key collider signature is that in any given decay chain of superpartners, there will be an additional $Z$-boson or Higgs boson in the final state with respect to the usual MSSM decay chain, which will then decay to SM fermions. 
Additionally, the decays of the LVSP to the LHSP are quite prompt. However, if the conditions are adequate, it is possible that there will be a displaced vertex from the LVSP decay. The parameter space of interest for displaced vertices is shown in Fig.~\ref{DispV.FIG}.

We have chosen the particular lengths shown from considerations of current searches at ATLAS and CMS, as well as the detector sizes. The CMS collaboration has published a search for displaced vertices in 8 TeV data where the minimal distance between two vertices $d_{vv}$ is required to be $600 \mu$m $\leq d_{vv} \leq 50$ mm \cite{Khachatryan:2016unx}. The ATLAS collaboration has published a search using 8 TeV data with a minimal distance between two vertices of $2$mm $\leq d_{vv} \leq 300$ mm \cite{Aad:2015rba}. Neither of these searches currently places interesting limits on the parameter space in which we are interested. However, this has motivated our choice of length scales shown in Fig.~\ref{DispV.FIG}, as well as displaying~1m and the approximate radius of both detectors of~10m. 
We see that for the benchmark value of $\epsilon =10^{-3}$, if the mass difference between the LVSP and the LHSP is $30 \lesssim \delta m \lesssim 70$ GeV, the displaced vertex from the LVSP decay could be detectable at the LHC.

To summarise, if R-parity is conserved, every superpartner produced in a collider will eventually decay to the LVSP. The LVSP will then decay into the LHSP and either a Higgs or $Z$ boson, which will themselves decay to SM fermions. Therefore, the most robust detector signal of LVSP decay would be seeing more particles in the final state than expected in a usual MSSM decay chain. Additionally, there is the possibility of observing displaced vertices if the mass difference between the LVSP and LHSP is small.

\subsubsection{Decay of the hidden sector gaugino into light hidden sector fermions}

In the previous subsection we explained how the LVSP decays into a hidden sector gaugino. However, this hidden sector gaugino is quite likely not the dark matter, which is instead composed of the light chiral fermions with masses $\mathcal{O}$(MeV). The exact nature of the light chiral fermion depends on the model-specific UV description of the hidden sector. An example of an $SU(5)$ model that gives light chiral fermions is presented in Appendix~\ref{Weakly}. For the purposes of calculating the decay of the hidden sector gaugino into the light chiral fermion, however, we can ignore the specifics of the UV completion, and consider a simple effective theory consisting of only the gaugino $\tilde{Z}'$ and the chiral superfields $\mathbf{X},~  {\mathbf{X}}',~\mathbf{X}''$ containing the chiral fermions $X,~X',~X''$ and their scalar partners $\tilde{X},~\tilde{X}',~\tilde{X}''$. The interaction part of the Lagrangian can then be written as:
\begin{align}
\mathcal{L}_{\rm int.} \supset -i\sqrt{2}g_H \left(\tilde{X}^\dagger Q_X \tilde{Z}' X - \overline{X}Q_X \overline{\tilde{Z}'} \tilde{X} \right) - \left(y_X \tilde{X} X' X'' + h.c.\right) \ ,
\end{align}
where $g_H$ is the hidden sector $U(1)$ gauge coupling, $Q_X$ is the charge and $y_X$ is the appropriate Yukawa coupling, which depends on the UV model considered.

\begin{figure}[H]
\centering
\subfloat[]{
\includegraphics[scale=0.6,valign=c]{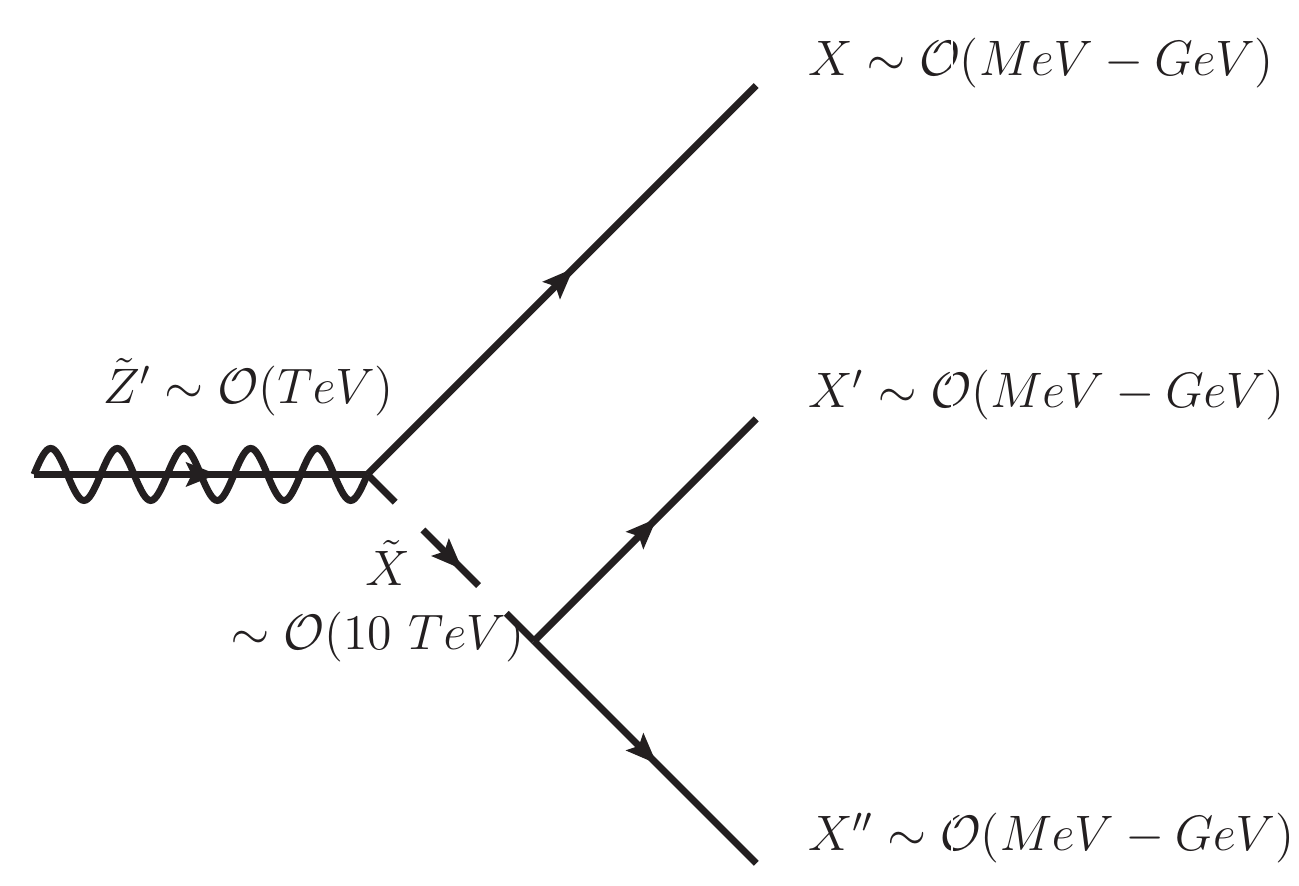}
\vphantom{\includegraphics[scale=0.6,valign=c]{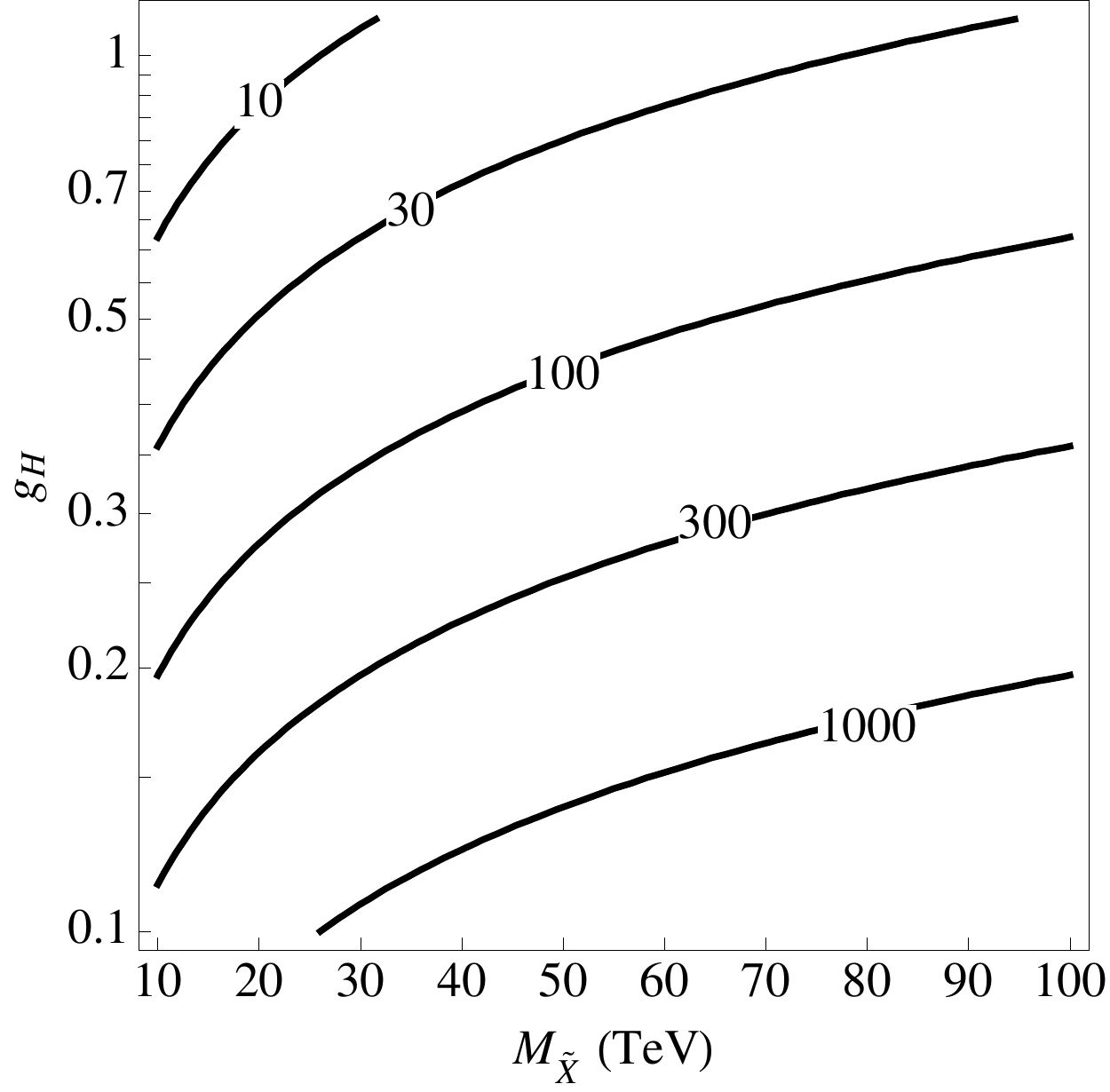}}
}
\subfloat[]{\includegraphics[scale=0.6,valign=c]{HGDecayTime.pdf}}
\caption{\\ \textbf{(a)} A diagram showing the potential decay of a hidden sector gaugino to hidden sector chiral fermions, including the dark matter candidate $X$.\\
\textbf{(b)} Contours showing the decay time (in seconds) of the hidden sector gaugino $\tilde{Z}'$ to three light hidden sector chiral fermions $f \equiv X,~X',~X''$, through a scalar mediator $\tilde{X}$, shown as a function of both the hidden sector gauge coupling $g_H$, and the mass of $\tilde{X}$. The mediator mass is expected to be of order $m_{3/2}$, and therefore in the range of 10-100 TeV. The gauge coupling range we have chosen so that it includes the couplings we found for the $SU(5)$ example in the Appendix. We have set the Yukawa coupling of the chiral fermions such that $M_f = 1$ MeV. We have taken the mass of the hidden gaugino to be $M_{\tilde{Z}'} \sim m_{3/2}/16\pi^2$, as could be expected for example, from anomaly mediation~\cite{Randall:1998uk, Giudice:1998xp,Gaillard:1999yb,Nelson:2002fk}.}
\label{HGDecay.FIG}
\end{figure}

The decay width of the hidden gaugino into chiral fermions is then
\begin{align}
\Gamma_{\tilde{Z}'} \sim \frac{1}{192\pi^3}y_X^2 (g_H Q_X)^2 \frac{M_{\tilde{Z}'}^5}{M_{\tilde{X}}^4}
\end{align}
in the limit where $M_X \ll M_{\tilde{Z}'}$, which is valid since we expect $M_{\tilde{Z}'}\sim \mathcal{O}$(TeV), while $1\text{ MeV} \lesssim M_X \lesssim 100$ GeV. As a result, we can see from Fig. \ref{HGDecay.FIG} that we expect the decay of the hidden sector gaugino to be sufficiently prompt. We define sufficiently prompt here to be such that the hidden sector gaugino lifetime is not greater than the age of the universe.

\section{Phenomenological Consequences}\label{Detection.SEC}

Having established in the previous section that the MSSM LVSP can decay into a hidden sector LHSP, which will then decay sufficiently promptly into hidden sector fermions, we now consider the phenomenology of the various possible dark matter candidates. 

As discussed previously, the annihilation cross-section which sets the relic abundance depends on hidden sector particles and their couplings only. However, interactions with the visible sector depend on the specifics of the kinetic mixing portal. Therefore, experimental signatures of these models will typically be suppressed in two ways. The first is by $\epsilon^2$ insertions wherever $Z-Z'$ mixing occurs. Additionally, there is a suppression of both scattering and Drell-Yan production of dark matter by the $Z'$ mass. Thus there will be factors of $\epsilon^2 \left(M_{Z}/M_{Z^\prime}\right)^4$ compared to usual visible sector WIMP models. Given that $\epsilon \lesssim 10^{-3}$ is expected in string/M-theory models~\cite{Dienes:1996zr,Bullimore:2010aj,Goodsell:2010ie,Goodsell:2011wn}, and assuming that $M_{Z^\prime} \sim (g_H v_H)$, the ability to detect dark matter in a hidden sector interacting with the visible sector will depend strongly on $g_H$ and $v_H$. In what follows, we will treat both of these as totally free parameters. We consider some variation of $g_H$ in Fig.~\ref{DirectDetection.FIG}, as it enters the dark matter-SM fermion scattering cross-section calculation in both the numerator and the denominator, and allow the value of $v_H$ to be set by choosing a value of $M_{Z'}$.

\subsection{Direct Detection in Nucleon Recoils}

There have been many attempts to detect dark matter through scattering off nucleons in materials. The strongest current limits have been set by LUX \cite{Akerib:2016vxi} for dark matter masses above 5 GeV, by CDMSlite \cite{Agnese:2015nto} for dark matter masses between 2 and 5 GeV, and finally by CRESST-II \cite{Angloher:2015ewa} for dark matter masses between 0.5 and 2 GeV. In the future, LZ is expected to increase the limits on $\sigma_{X-N}$ in the same mass range as LUX \cite{Akerib:2015cja}, while SuperCDMS will strengthen limits at a slightly lower mass range, down to about $M_X \sim 0.4$ GeV \cite{Agnese:2016cpb}, thus improving on CDMSlite and CRESST-II. The proposed DARWIN experiment would also probe the $M_X > 5$ GeV region \cite{Aalbers:2016jon}.

In order to compute the dark matter-nucleon scattering cross-section via a kinetically mixed $Z'$, we use the results of~\cite{Babu:1997st, Frandsen:2012rk} for the mixing between the SM $Z$ and the $Z'$. The effective Lagrangian is
\beq
\Lagr_{\rm eff} = C_{f}^V \bar{X}\gamma^\mu X \bar{f}\gamma_\mu f \ ,
\eeq
where the effective coupling $C_f^V$ is given by
\beq
C_f^V = \frac{g^{Z'}_{X}g^{Z'}_{f}}{M_{Z'}^2}+\frac{g^Z_{X}g^Z_{f}}{M_{Z}^2}\ ,
\eeq
where the couplings $g^{Z'}_{X}=g_H$ and $g^Z_{f}$ exist at tree-level, but the couplings $g^{Z'}_{f}$ and $g^Z_{X}$ are induced only via the kinetic mixing. The coupling to protons and neutrons is then given by
\beq
C^V_p = 2C_u^V + C_d^V,~~ C^V_n = 2C_d^V + C_u^V \ .
\eeq
When there is no mass mixing between the $Z$ and $Z'$, the coupling to neutrons is vastly subdominant to the coupling to protons, and can therefore be neglected. Explicitly, the coupling to protons is
\beq
C^V_p \simeq    \epsilon g_Y g_H \frac{\left(1-4s_w^2 \right)}{M_{Z'}^2} \ ,
\eeq
in the limit of small kinetic mixing parameter $\epsilon$. The coupling $g_Y$ is the usual hypercharge gauge coupling, and we have abbreviated $s_w= \sin\theta_w$. For the full expressions, see \cite{Frandsen:2012rk}.

Since the spin-independent dark matter-nucleon scattering cross-section is given by
\beq
\sigma_{X-N} = \frac{\mu_{XN}^2 ({C^V_p}^2 + {C^V_n}^2)}{\pi} \ ,
\eeq
where $\mu_{XN}$ is the reduced mass of the dark matter candidate and the nucleon. In the pure kinetic mixing case with small mixing parameter $\epsilon$, this simplifies to
\beq
\sigma_{X-N} \simeq \frac{\epsilon^2 g_Y^2 g_H^2 \mu_{XN}^2}{ \pi } \frac{\left(1-4s_w^2 \right)^2}{M_{Z'}^4} \ . 
\eeq
This expression accurately captures both the low- and high-$M_{Z'}$ regimes.

\subsection{Direct Detection in Electron Recoil Experiments}

When the dark matter candidate $X$ is very light, and the $Z'$ is significantly lighter than the $Z$-boson, dark matter interactions could be best probed by current/future dark matter-electron scattering experiments~\cite{Essig:2011nj, Graham:2012su, Essig:2012yx, Agnese:2013jaa, Essig:2015cda, Lee:2015qva, Hochberg:2015pha, Hochberg:2015fth,Hochberg:2016ntt, Alexander:2016aln}. The spin-independent dark matter-electron scattering cross-section is given by~\cite{Essig:2011nj}
\beq
\sigma_{X-e} = \frac{4 \mu_{Xe}^2 \alpha_{\rm em} g_H^2 c_w^2\epsilon^2}{M_{Z'}^4} \ ,
\eeq
for the models we consider, where $M_{Z'} \gg \alpha_{\rm em} m_e$ \footnote{In this expression there is an additional factor of $c_w^2$ due to a difference in the normalisation of $\epsilon$ between this work and~\cite{Essig:2011nj}.}. In the above equation, $\mu_{X e}$ is the reduced mass of the dark matter and electron, $\alpha_{\rm em}$ is the usual EM coupling. Since we are in the heavy $Z'$ regime, the dark matter form factor~\cite{Essig:2011nj} is
\beq
F_{dm}(q) = \frac{M_{Z'}^2 + \alpha_{\rm em}^2 m_e^2}{M_{Z'}^2 + q^2} \simeq 1 \ ,
\eeq
for our choice of parameters.

While Xenon10 is not sensitive to the regions of interest \cite{Essig:2017kqs}, future experiments could be sensitive. In particular, scattering off Silicon imposes the strongest future limits \cite{Essig:2015cda,Alexander:2016aln} in the region of interest. If the dark matter were to be lighter than 1 MeV, with a light mediator $Z'$, corresponding to when the relic density is set by freeze-out during radiation domination (given by Eq.~(\ref{FOrad.EQ})), then superconductors could be the strongest probe~\cite{Hochberg:2015fth, Hochberg:2015pha}. 

\subsection{Direct Collider Searches and Electroweak Precision Constraints on $Z'$ Mediated Dark Matter}

Direct collider searches and electroweak precision constraints are primarily applicable to constraining a combination of $\epsilon$ and $M_{Z'}$, which indirectly constrains the allowed parameter space for our dark matter candidate to live in. The direct detection cross-section is proportional to $\epsilon^2$, while the relic density calculation is independent of $\epsilon$. Both processes depend the same way on $M_{Z'}$. Therefore, we would like to use direct collider and precision constraints to consider the maximal value of $\epsilon$ to allow for maximal coverage in direct detection experiments. 

As long as $\epsilon \leq 10^{-3}$, a $Z'$ that kinetically mixes with the $Z$ will not be probed directly at the LHC \cite{Curtin:2014cca} for $M_{Z'} \geq 10$ GeV. The strongest limits are from direct Drell-Yan production of the $Z'$, and can be sensitive to  $\epsilon \gtrsim 9 \times 10^{-4}$ at HL-LHC for $10 \text{ GeV}< M_{Z'} < M_Z$, and up to $\epsilon \gtrsim 2 \times 10^{-3}$ for $M_{Z'} > M_Z$ \cite{Curtin:2014cca}. For $1\,{\rm GeV} < M_{Z'} \leq 10\,{\rm GeV}$, BaBar places the strongest limit, with $\epsilon \lesssim 7 \times 10^{-4}$ for $1\,{\rm GeV} < M_{Z'} \leq 10\,{\rm GeV}$~\cite{Lees:2014xha}. Therefore, the parameter space we consider above is safe from all direct collider constraints. 

Additionally, electroweak precision tests are insensitive to the choices of parameters we have made. The most stringent electroweak precision constraint is on the $Z$-pole, and could eventually impose $\epsilon \leq 7 \times 10^{-4}$ at ILC/GigaZ \cite{Curtin:2014cca}. For $10 \text{ GeV}< M_{Z'} < M_Z$, electroweak precision measurements can constrain $\epsilon \leq 3.5 \times 10^{-3}$ at ILC/GigaZ. We expect that a potential FCC$-ee$ or TeraZ machine would be able to probe $\epsilon$ even further. The eventual HL-LHC with 3000 fb$^{-1}$ can only constrain $\epsilon \leq 2 \times 10^{-3},~1.3\times10^{-2}$ on the $Z$-pole and below, respectively.

Note that beam dump experiments such as LSND \cite{deNiverville:2011it} and E137 \cite{Batell:2014mga}, which typically constrain kinetically mixed $Z'$-mediated dark matter candidates quite strongly \cite{deNiverville:2011it, Batell:2014mga, Alexander:2016aln}, do not constrain our parameter space. This is due to the relic density in our analysis being set entirely by hidden sector processes, whereas in most analyses it is set by $HS \leftrightarrow SM$ processes. Therefore the cross-sections for setting the dark matter relic density, and for interacting with the Standard Model, do not have the same parametric dependence in our analysis, whereas in other analyses they are strongly correlated.

\begin{figure}[t]
\centering
\includegraphics[scale=0.85]{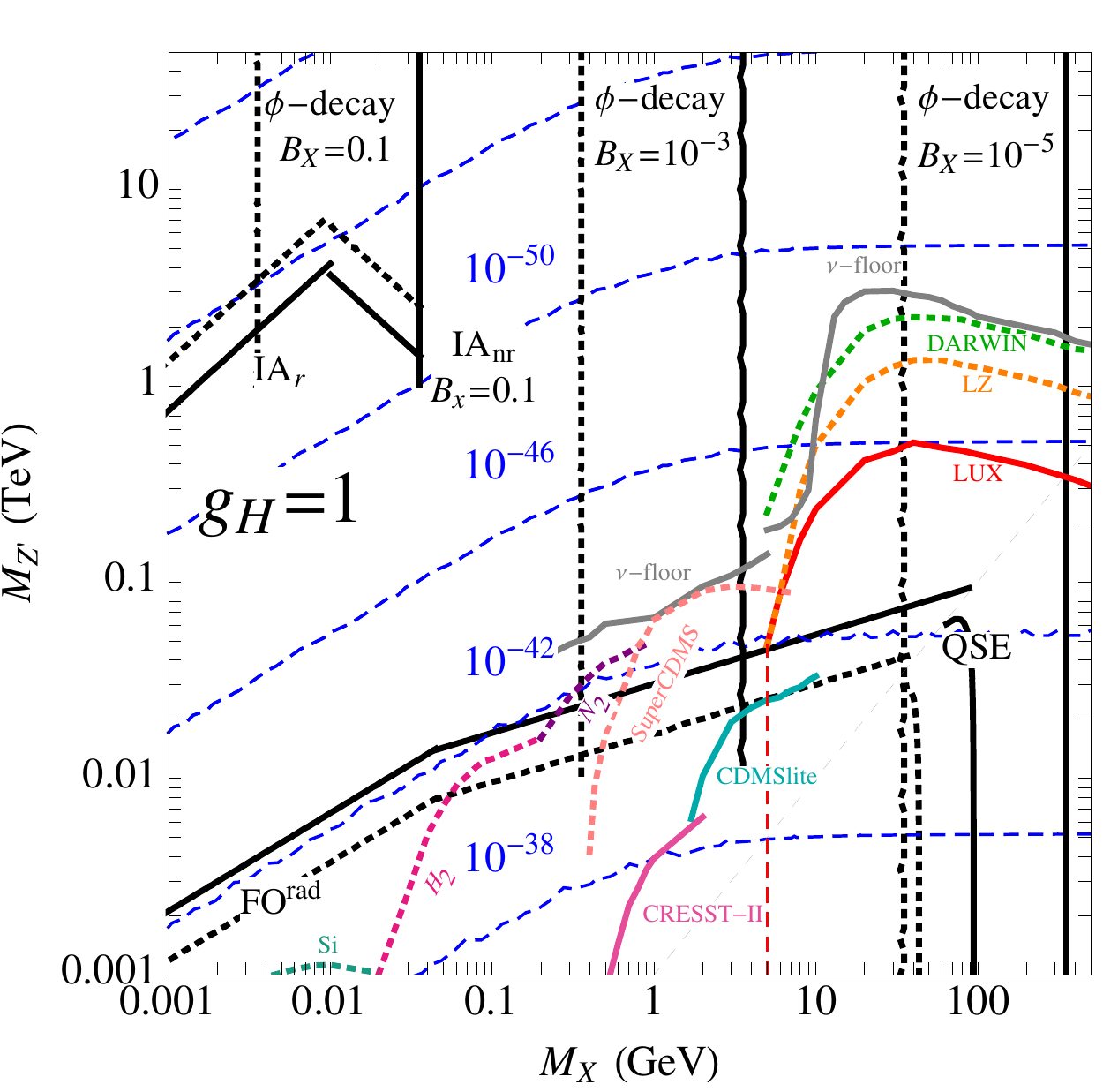}
\caption{Current exclusions and future sensitivity in the hidden sector dark matter mass $M_X$, $Z'$ mediator mass $M_{Z'}$ plane, with the hidden sector $U(1)$ gauge coupling $g_H=1$. The dashed blue contours show $\sigma_{X-N}^{SI}$ for Xenon, in units of (cm$^2$). Below the red contour is currently excluded by LUX~\cite{Akerib:2016vxi}. The future limits from LZ are shown in dotted orange~\cite{Akerib:2015cja}, while those from Darwin are shown in green~\cite{Aalbers:2016jon}. The current limits from CDMSlite are shown in dark cyan~\cite{Agnese:2015nto}. The current limits from CRESST-II~\cite{Angloher:2015ewa} are shown in pink. The light pink dotted contour shows the projected sensitivity of the Si and Ge HV detectors of SuperCDMS SNOLAB~\cite{Agnese:2016cpb}. The dashed contours in the lower left indicate future sensitivity of potential $H_2$ (pink), $N_2$ (purple) dissociation experiments~\cite{Essig:2016crl}. The dark green dotted line in the lower left corner is the future sensitivity of dark matter scattering off electrons in Si semiconductors\cite{Essig:2015cda}. Above the grey solid line is below the neutrino floor as reported in~\cite{Angloher:2015ewa} and~\cite{Billard:2013qya}. The black solid (dotted) lines indicate where the dark matter relic density is $\Omega_{dm} h^2 = 0.12~(0.012)$. Labels indicate what mechanism is setting the relic density. The IA$_{nr}$ mechanism is only displayed for $B_X=0.1$. For smaller $B_X$ the IA$_{nr}$ validity region extends linearly up to a maximum of $M_X \sim 0.3$ GeV. See Table \ref{Mech.TAB} for definitions for each mechanism. $\phi$ is the lightest modulus. The kinetic mixing parameter is set to $\epsilon=10^{-3}$. The light dashed line corresponds to $M_X = M_{Z'}$. All values of $M_X$ shown correspond to some viable dark matter candidate.}
\label{DirectDetection.FIG}
\end{figure}

\begin{figure}[t]
\centering
\includegraphics[scale=0.85]{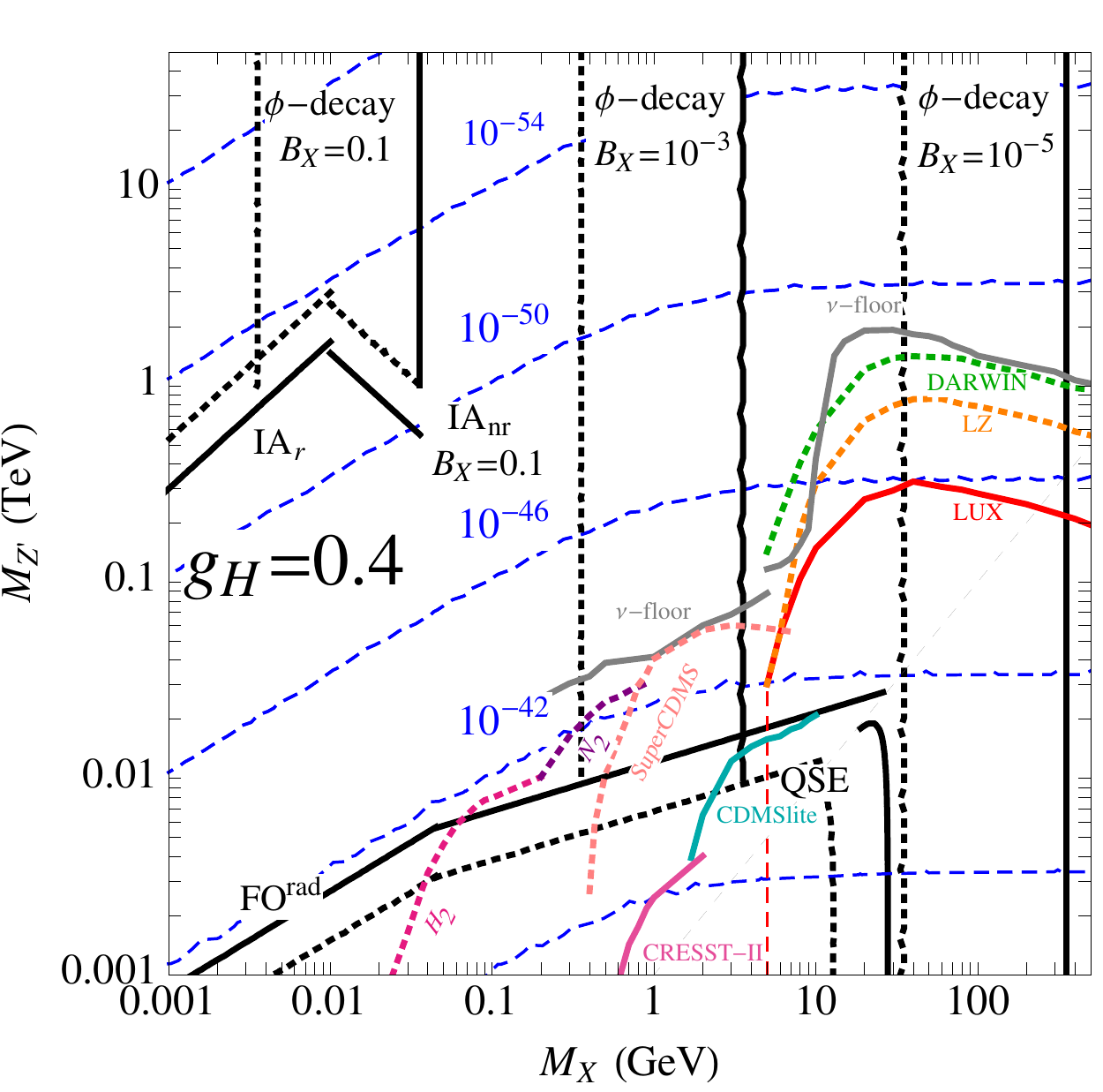}
\caption{As in Fig. \ref{DirectDetection.FIG}, with $g_H = 0.4$.}
\label{DirectDetection2.FIG}
\end{figure}

For the choices of parameters in Eq.~(\ref{BM.EQ}), we see from Figs.~\ref{DirectDetection.FIG}, \ref{DirectDetection2.FIG} that the QSE solution in Eq.~(\ref{QSE.EQ}) is ruled out by LUX for $M_X\gtrsim5$ GeV regardless of our choice of $g_H$. In the future, SuperCDMS will be able to rule out the QSE solution for $M_X \gtrsim 500~ (600)$ MeV~\cite{Agnese:2016cpb} for $g_H=0.4~ (1)$. Note that when $M_X > M_{Z'}$, the QSE solution no longer depends strongly on the mediator mass, as the cross-section now scales as $1/M_X^2$.

No planned experiment will be able to probe the FO$^{rad}$ mechanism in Eq.~(\ref{FOrad.EQ}) unless $g_H \gg 1$, in which case semiconductors could present a possible probe~\cite{Essig:2015cda}. Potentially superconductors could also probe ultra-light dark matter for this mechanism~\cite{Hochberg:2015fth, Hochberg:2015pha}. However, this would correspond to $M_{Z'} \ll 1$ GeV, which we do not present here, due to the extreme tuning of $v_H$ relative to $m_{3/2}$ this would require given an $\Order(1)$ coupling $g_H$.

The modulus decay mechanism can only be probed for $B_X \lesssim 10^{-3}$. If $B_X=10^{-3}$, the current strongest limit is from CDMSlite, which requires $M_{Z'} \gtrsim 20~(12)$ GeV for $g_H=1~(0.4)$. In the future, SuperCDMS will be sensitive to $M_{Z'} \sim 100~(60)$ GeV. This will be near the neutrino floor, which makes detection above $M_{Z'} \simeq 110~(70)$ GeV more difficult.
 
If $B_X=10^{-5}$, the current strongest limit is from LUX, which constrains $M_{Z'} \gtrsim 360~(230)$ GeV for $g_H=1~(0.4)$. In the future, LZ will be sensitive to $M_{Z'} \sim 1000~(640)$ GeV. The possible DARWIN experiment would be sensitive to $M_{Z'} \sim 2~(1)$ TeV. The neutrino floor lies at $M_{Z'} \sim 1.9~(1.1)$ TeV. 

If $B_X >10^{-3}$, the region which gives the correct relic density will lie beyond the projected sensitivity of all experiments shown here, and also beyond the neutrino floor, making detection impossible in the near future.

The IA$_r$ and IA$_{nr}$ mechanisms are also impossible to probe experimentally in the near future. Given the vast increase in sensitivity required to reach these regions of parameter space, between 12 and 16 orders of magnitude, it is highly unlikely that these regions will be testable in direct detection experiments. However, since this corresponds to relatively large $Z'$ masses, it is possible that these could be probed eventually at colliders~\cite{Curtin:2014cca} if $\epsilon \gtrsim \mathcal{O}(\mathrm{few} \times 10^{-4})$.

\section{Conclusions}\label{conclusions}

\begin{table}[t]
\resizebox{\textwidth}{!}{
\centering
\begin{tabular}{c | c | c | c | c}
\textbf{Mechanism} & \textbf{Candidate} & \textbf{Mediator} & \textbf{Current status} & \textbf{Future status} \\
\hline
&&&&\\
QSE (Eq. (\ref{QSE.EQ})) & \textit{chiral fermion, (light) gaugino} & \textit{gauge boson} & LUX\cite{Akerib:2016vxi} &SuperCDMS \cite{Agnese:2016cpb} \\
& $0.1 \lesssim M_X \lesssim 100$ GeV & $10 \lesssim M_{Z'} \lesssim 100$ GeV & $M_X \lesssim 5$ GeV & $M_X \lesssim 500$ MeV  \\
FO$^{rad}$ (Eq. (\ref{FOrad.EQ})) & \textit{chiral fermion}  & \textit{gauge boson} & -- & -- \\
& $M_X \lesssim 100$ MeV & $M_{Z'} \lesssim 10$ GeV & & \\
&&&&\\
\hline
&&&&\\
Modulus decay (Eq. (\ref{modDec.EQ})) & \textit{chiral fermion} & \textit{gauge boson, scalar} & -- & -- \\
$B_X\sim 10^{-1}$ & $100 \lesssim M_X \lesssim 500$ MeV  & $1 \lesssim M_{M} \lesssim 100$ TeV & & \\
 & \textit{chiral fermion} & \textit{gauge boson, scalar} & CDMSlite \cite{Agnese:2015nto}& SuperCDMS\cite{Agnese:2016cpb}  \\
$B_X\sim 10^{-3}$ & $0.5 \lesssim M_X \lesssim 5$ GeV  & $10 \lesssim M_{M} \lesssim 10^6$ GeV & $M_{Z'}\gtrsim \Order(10)$ GeV & $M_{Z'}\gtrsim \Order(100)$ GeV\\
& \textit{chiral fermion, gaugino} & \textit{gauge boson, scalar} & LUX\cite{Akerib:2016vxi} & LZ\cite{Akerib:2015cja}, DARWIN \cite{Aalbers:2016jon}  \\
$B_X\sim 10^{-5}$ & $50 \lesssim M_X \lesssim 500$ GeV  & $10 \lesssim M_{M} \lesssim 10^6$ GeV & $M_{Z'}\gtrsim \Order(300)$ GeV & $M_{Z'}\gtrsim \Order(1)$ TeV \\
&&&&\\
\hline
&&&&\\
IA$_{nr}$ (Eq. (\ref{IANR.EQ})) & \textit{chiral fermion}  & \textit{gauge boson} & -- & -- \\
 & $10 \lesssim M_X \lesssim 100$ MeV & $100 \lesssim M_{Z'} \lesssim 10^4$ GeV & &  \\ 
IA$_r$ (Eq. (\ref{IAR.EQ})) & \textit{chiral fermion}  & \textit{gauge boson} & -- & -- \\
 & $M_X \lesssim 5$ MeV & $10 \lesssim M_{Z'} \lesssim 5000$ GeV & &  \\ 
 &&&&\\
 \hline
\end{tabular}
}
\caption{Table summarising the various relic density production mechanisms, as discussed in Section \ref{cosmo}, and the current allowed parameter space, as discussed in Section \ref{Detection.SEC}. The kinetic mixing parameter has been set to $\epsilon = 10^{-3}$.}
\label{Summary.TAB}
\end{table}

In the context of a string/$M$-theoretic UV completion of the visible sector MSSM, we have argued that hidden sector dark matter is well motivated~\cite{Acharya:2016fge}. As such, it is of great interest to consider the possible categories into which hidden sector dark matter can fall, based on the production mechanisms first studied in~\cite{Kane:2015qea}. We have argued here that it is highly likely that in a compactified string/$M$-theory, if dark matter is weakly coupled, it must be a fermion. Additionally, in most of the viable parameter space, it should be a relatively light chiral fermion with mass in the MeV to GeV range. In some parts of the parameter space, dark matter could be a hidden sector gaugino in the 100 GeV mass range. String/$M$-theory models could also contain strongly coupled hidden sectors, as well as axions, which we have not discussed here, but are also generically present \cite{Svrcek:2006yi}. Combining the results of \cite{Acharya:2016fge},  \cite{Halverson:2016nfq, Acharya:2017szw} and \cite{Svrcek:2006yi, Arvanitaki:2009fg}, the possibility emerges that the Universe has several types of non-thermally produced dark matter particles, each with very different properties and interactions.

We have expanded on previous analyses to discuss exactly why and how the visible sector LSP decays into the hidden sector lightest gaugino (hidden LSP), and what the most distinctive experimental features of such a decay might be, in Section \ref{decaypheno}. We have also argued that if the lightest hidden gaugino is not actually the lightest hidden sector state, it can then decay into light fermions in the hidden sector, which can be the dark matter. Thus the chain of decays from visible LSP $\to$ hidden gaugino $\to$ dark matter is theoretically achieved. We present in an appendix a string-motivated hidden sector which could give rise to a weakly coupled dark matter candidate.

Given the viable dark matter production mechanisms discussed in \cite{Kane:2015qea}, we have presented how direct detection experiments and direct collider searches constrain the parameter space. Our findings are summarised in Table \ref{Summary.TAB}. We find that two production mechanisms for hidden sector dark matter have direct detection signatures that are observable, either currently, or at planned experiments. These are the ``Quasi-static equilibrium" solution, which is the generalisation of the ``non-thermal WIMP miracle", and modulus decay. We also find, unfortunately, that the remaining three mechanisms: freeze-out during radiation domination, relativistic inverse annihilation and non-relativistic inverse annihilation (inverse annihilation is when dark radiation annihilates back into dark matter), are very difficult to search for at direct detection experiments. A summary of the current and future experimental tests of these mechanisms is shown in Table \ref{Summary.TAB}. The two inverse annihilation mechanisms might be probed at future colliders.

\subsubsection*{Acknowledgements}

We thank John Ellis, Gordan Krnjaic, Bibhushan Shakya, Aaron Pierce and Yue Zhao for useful discussions. We thank Aaron Pierce particularly for detailed comments on the manuscript. SARE thanks the DESY theory group for their hospitality while this work was being completed. The work of BSA was supported by the STFC Grant ST/L000326/1. This work was supported by a grant from the Simons Foundation (\#488569, Bobby Acharya). The work of SARE and GLK is supported in part by the U.S. Department of Energy, Office of Science, under grant DE-SC0007859. The work of SARE is also supported in part by a Rackham research grant. The work of BN is supported in part by the National Science Foundation, under grant PHY-1620575. The work of MJP is supported in part by STFC.

\appendix
\section{An $SU(5)$ Hidden Sector Model}\label{Weakly}
\label{SU5.APP}

In this appendix we give an explicit example of a weakly-coupled hidden sector model which can arise from compactified string/M-theories. More specifically, we will consider a non-Abelian hidden sector which can naturally arise in M-theory~\cite{Acharya:2000gb, Witten:2001bf, Friedmann:2002ty}. Motivated by visible sector $SU(5)$ GUTs, we consider the possibility of an $SU(5)$ hidden sector which will be broken either at tree-level or by certain fields acquiring expectation values. In order to have anomaly cancellation, we take our field content to consist of $\mathbf{10}_H,~\mathbf{5}_H,~\mathbf{\bar{5}}$. In our notation, the subscript $H$ denotes a field which will obtain a \emph{vev}, $v_H$, through spontaneous symmetry breaking. 

In order for sizeable kinetic mixing to occur between the visible and hidden sectors, we assume that Wilson line breaking occurs in both sectors. However, we will show that the hidden gauge symmetries can also undergo radiative symmetry breaking, giving rise to a low-scale vacuum expectation value. We also assume that gaugino masses in the hidden sector are suppressed relative to the gravitino mass $m_{3/2}$, as they are expected to experience SUSY breaking in the same way as the visible sector. We do not attempt to calculate the mass of the $Z'$ in this model, but assume that we can adjust $v_H$ in order to achieve the range of masses we consider in Section~\ref{Detection.SEC}.
Finally, since a hierarchy in the Yukawa couplings will be necessary to ensure that there is a light chiral dark matter candidate, we will consider how to achieve such a hierarchy.

\subsection*{Starting from $E_8$}

In order to get the $SU(5)$ singularities where the chiral fermions are found, we start from an $E_8$ singularity. This is motivated by considering the $M$-theory limit of the heterotic string, where the $E_8$ is broken to $SU(5) \times SU(5)'$. The decomposition of the $\mathbf{248}$ of $E_8$ into $SU(5) \times SU(5)'$ is the following
\begin{align}
\mathbf{248} = (\mathbf{1},\mathbf{24}) + (\mathbf{24},\mathbf{1}) + (\mathbf{5},\mathbf{\overline{10}}) + (\mathbf{\overline{5}},\mathbf{10}) + (\mathbf{10},\mathbf{5}) + (\mathbf{\overline{10}},\mathbf{\overline{5}}) \, .
\end{align}
Thus an $E_8$ superpotential
\begin{align}
W = \mathbf{248}~\mathbf{248}~\mathbf{248} \, ,
\end{align}
will give an $SU(5)$ superpotential that is consistent with all the charges under both $SU(5)$ and $SU(5)'$, that contains
\begin{align}
W \supset \mathbf{10}~\mathbf{10}~\mathbf{5} + \mathbf{10}~\mathbf{\overline{5}}~\mathbf{\overline{5}}\, ,
\end{align}
where each $SU(5)$ superfield is localized on a different singularity. This allows for there to be a hierarchy in the Yukawa couplings of the two terms in the superpotential.

\subsection*{The effective model}
We are then in a position where we can write the relevant terms of the superpotential as:
\begin{align}
W = \lambda \mathbf{10}_H ~\mathbf{10}_H ~\mathbf{5}_H  
+ \lambda^{\prime \prime} ~\mathbf{10}_H ~\mathbf{\overline{5}} ~\mathbf{\overline{5}}
\label{EfSP.EQ}
\end{align}
where this superpotential can be obtained from the $E_8$ superpotential as explained above.

In this superpotential we can assume $\lambda \sim \mathcal{O}(1)$ and $\lambda^{\prime \prime}\ll \lambda$, as motivated above. We subsequently write the Higgs part of the scalar potential as 
\begin{align}
V_H = m_H^2 ({\chi^{\dagger}}^{ij} \chi_{ij} + {\phi^{\dagger}}^m \phi_m) + \left|\frac{\partial W}{\partial \chi_{ij}}\right|^2 + \left|\frac{\partial W}{\partial \phi_{m}}\right|^2 + \frac{g^2}{2}\left( \sum_a \left[ {\phi^{\dagger}}^i {T^a}_i^j\phi_j + \frac{1}{2} {\chi^\dagger}^{ij} \left( {T^a}_i^k \chi_{kj} +{T^a}_j^l \chi_{il} \right)\right] \right)^2
\end{align}
where the $\mathbf{10}_H$ is written as $\chi_{ij}$ and the $\mathbf{5}_H$ is written as $\phi_m$, with $i,j,m = 1,\ldots, 5$. 

For tree-level symmetry breaking, we look for solutions to the fifteen coupled equations:
\begin{align}
\partial_{\chi_{ij}}V = 0,~~\partial_{\phi_m}V =0
\end{align}
We find that in general the form of the vacuum expectation values acquired by the $\mathbf{10}_H$ and the $\mathbf{5}_H$ can be written as
\begin{align}
\langle\mathbf{10}\rangle = 
\bmat
0 & a & 0 & 0 & 0 \\
-a & 0 & 0 & 0 & 0 \\
0 & 0 & 0 & b & 0 \\
0 & 0 & -b & 0 & 0 \\
0 & 0 & 0 & 0 & 0
\emat
,~~~~ \langle \mathbf{5}' \rangle = \bmat c \\ d \\ e \\ f \\ g \emat
\end{align}
There are then several solutions, two of which correspond to the lowest vacuum state after symmetry breaking, with an unbroken $SU(2) \times U(1)$ symmetry. These two solutions are
\begin{align}
\nonumber a = 0, ~~ b \neq 0, ~~ c \neq 0, ~~ d= 0, ~~ e =0,~~ f=0,~~g\neq 0 \\
a \neq 0, ~~ b \neq 0, ~~ c \neq 0, ~~ d= 0, ~~ e =0,~~ f=0,~~g = 0
\end{align}
The leftover $U(1)$ can then mix via the kinetic mixing portal with the visible sector.

We also investigate the possibility of radiative symmetry breaking here. The RGEs, setting the scalar masses degenerate, are found to be 
\begin{align}
\beta_{m_H^2} &= \frac{1}{16\pi^2} \left( 18 \lambda^2m_H^2 + 6\tilde{A}^2\right) \ , \\
\beta_{\tilde{A}} &= \frac{1}{16\pi^2} \left( 27 \lambda^2 \tilde{A} \right) + \frac{1}{5\pi^2} \left( -6 g^2 \tilde{A}\right) \ ,\\
\beta_{\lambda} &= \frac{1}{16\pi^2} \left( 9 \lambda^3 \right)+\frac{1}{5\pi^2} \left( -6 g^2\lambda \right) \ ,\\
\beta_{g} &= \frac{1}{4\pi^2} \left( -3g^3\right) \ ,
\end{align}
where $\tilde{A} = A \lambda$, with $A$ the trilinear term,
which when solved, allows us to find regions of parameter space where the Higgs mass-squared is driven negative. We find that a fairly large trilinear is required, but relatively small gauge and Yukawa couplings are sufficient to radiatively break the symmetry.

\begin{figure}[H]
\centering
\includegraphics[scale=0.7]{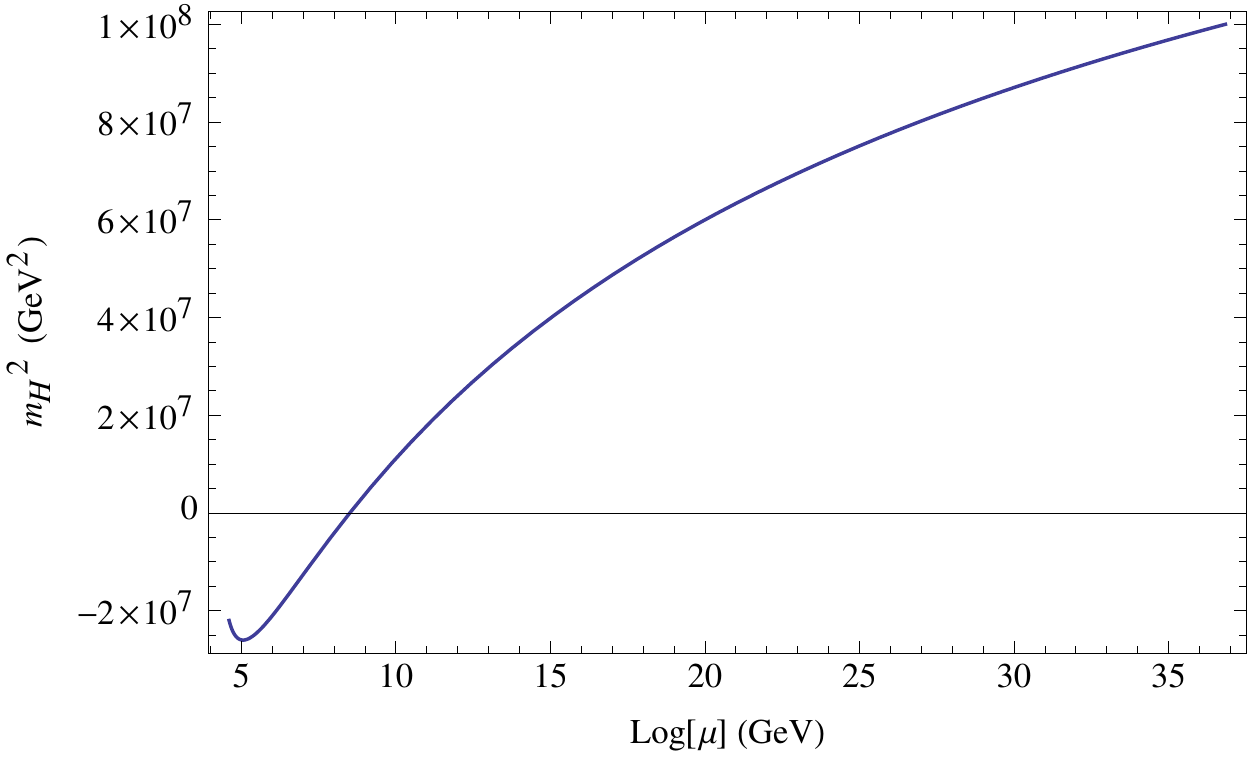}
\caption{Plot showing the mass-squared going tachyonic for the $SU(5)$ model with $\lambda = 0.2$, $g= 0.45$, $A= \sqrt{8} m_H$ and $m_H(m_{GUT}) = 10$ TeV. }
\label{SU5RGE.FIG}
\end{figure}

The HS dark matter candidate in this toy model would be the lightest stable fermionic component of the $\mathbf{\overline{5}}$  whose mass is of order $M_{X} \sim \lambda^{\prime \prime} \left< \mathbf{10}_H\right>$ (see Table \ref{Summary.TAB} for the conditions on $M_{X}$ for the various production mechanisms so that $X$ is a good dark matter candidate). Thus, we have shown that in this model, a chiral fermion whose mass is generated by the $SU(5)$ breaking provides a good DM candidate in non-thermal cosmological histories, with mass $\sim \mathcal{O}$(few) MeV - $\mathcal{O}$(few) GeV, thus covering the full range we have studied in this paper. Of note is that this hidden sector model contains other particles beyond that which is the dark matter. These could make up the dark radiation, for example.

\bibliographystyle{h-physrev}
\bibliography{NotesAug5}

\begin{thebibliography}{10}

\bibitem{Preskill:1982cy}
J.~Preskill, M.~B. Wise, and F.~Wilczek,
\newblock Phys. Lett. {\bf 120B}, 127 (1983).

\bibitem{Abbott:1982af}
L.~F. Abbott and P.~Sikivie,
\newblock Phys. Lett. {\bf 120B}, 133 (1983).

\bibitem{Dine:1982ah}
M.~Dine and W.~Fischler,
\newblock Phys. Lett. {\bf 120B}, 137 (1983).

\bibitem{Pagels:1981ke}
H.~Pagels and J.~R. Primack,
\newblock Phys. Rev. Lett. {\bf 48}, 223 (1982).

\bibitem{Weinberg:1982zq}
S.~Weinberg,
\newblock Phys. Rev. Lett. {\bf 48}, 1303 (1982).

\bibitem{Coughlan:1983ci}
G.~D. Coughlan, W.~Fischler, E.~W. Kolb, S.~Raby, and G.~G. Ross,
\newblock Phys. Lett. {\bf 131B}, 59 (1983).

\bibitem{deCarlos:1993wie}
B.~de~Carlos, J.~A. Casas, F.~Quevedo, and E.~Roulet,
\newblock Phys. Lett. {\bf B318}, 447 (1993), hep-ph/9308325.

\bibitem{Acharya:2016fge}
B.~S. Acharya, S.~A.~R. Ellis, G.~L. Kane, B.~D. Nelson, and M.~J. Perry,
\newblock Phys. Rev. Lett. {\bf 117}, 181802 (2016), 1604.05320.

\bibitem{Moroi:1999zb}
T.~Moroi and L.~Randall,
\newblock Nucl. Phys. {\bf B570}, 455 (2000), hep-ph/9906527.

\bibitem{Acharya:2008bk}
B.~S. Acharya {\em et~al.},
\newblock JHEP {\bf 06}, 064 (2008), 0804.0863.

\bibitem{Cohen:2013ama}
T.~Cohen, M.~Lisanti, A.~Pierce, and T.~R. Slatyer,
\newblock JCAP {\bf 1310}, 061 (2013), 1307.4082.

\bibitem{Fan:2013faa}
J.~Fan and M.~Reece,
\newblock JHEP {\bf 10}, 124 (2013), 1307.4400.

\bibitem{Blinov:2014nla}
N.~Blinov, J.~Kozaczuk, A.~Menon, and D.~E. Morrissey,
\newblock Phys. Rev. {\bf D91}, 035026 (2015), 1409.1222.

\bibitem{Denef:2004cf}
F.~Denef and M.~R. Douglas,
\newblock JHEP {\bf 03}, 061 (2005), hep-th/0411183.

\bibitem{Acharya:2010af}
B.~S. Acharya, G.~Kane, and E.~Kuflik,
\newblock Int. J. Mod. Phys. {\bf A29}, 1450073 (2014), 1006.3272.

\bibitem{Holdom:1985ag}
B.~Holdom,
\newblock Phys. Lett. {\bf B166}, 196 (1986).

\bibitem{Dienes:1996zr}
K.~R. Dienes, C.~F. Kolda, and J.~March-Russell,
\newblock Nucl. Phys. {\bf B492}, 104 (1997), hep-ph/9610479.

\bibitem{Halverson:2016nfq}
J.~Halverson, B.~D. Nelson, and F.~Ruehle,
\newblock Phys. Rev. {\bf D95}, 043527 (2017), 1609.02151.

\bibitem{Acharya:2017szw}
B.~S. Acharya, M.~Fairbairn, and E.~Hardy,
\newblock (2017), 1704.01804.

\bibitem{Dienes:2016vei}
K.~R. Dienes, F.~Huang, S.~Su, and B.~Thomas,
\newblock Phys. Rev. {\bf D95}, 043526 (2017), 1610.04112.

\bibitem{Soni:2016gzf}
A.~Soni and Y.~Zhang,
\newblock Phys. Rev. {\bf D93}, 115025 (2016), 1602.00714.

\bibitem{Soni:2017nlm}
A.~Soni, H.~Xiao, and Y.~Zhang,
\newblock (2017), 1704.02347.

\bibitem{Feldman:2007wj}
D.~Feldman, Z.~Liu, and P.~Nath,
\newblock Phys. Rev. {\bf D75}, 115001 (2007), hep-ph/0702123.

\bibitem{Pospelov:2007mp}
M.~Pospelov, A.~Ritz, and M.~B. Voloshin,
\newblock Phys. Lett. {\bf B662}, 53 (2008), 0711.4866.

\bibitem{Feng:2008mu}
J.~L. Feng, H.~Tu, and H.-B. Yu,
\newblock JCAP {\bf 0810}, 043 (2008), 0808.2318.

\bibitem{ArkaniHamed:2008qn}
N.~Arkani-Hamed, D.~P. Finkbeiner, T.~R. Slatyer, and N.~Weiner,
\newblock Phys. Rev. {\bf D79}, 015014 (2009), 0810.0713.

\bibitem{Pospelov:2008jd}
M.~Pospelov and A.~Ritz,
\newblock Phys. Lett. {\bf B671}, 391 (2009), 0810.1502.

\bibitem{Feng:2009mn}
J.~L. Feng, M.~Kaplinghat, H.~Tu, and H.-B. Yu,
\newblock JCAP {\bf 0907}, 004 (2009), 0905.3039.

\bibitem{Cohen:2010kn}
T.~Cohen, D.~J. Phalen, A.~Pierce, and K.~M. Zurek,
\newblock Phys. Rev. {\bf D82}, 056001 (2010), 1005.1655.

\bibitem{Kane:2015qea}
G.~L. Kane, P.~Kumar, B.~D. Nelson, and B.~Zheng,
\newblock (2015), 1502.05406.

\bibitem{Acharya:2009zt}
B.~S. Acharya, G.~Kane, S.~Watson, and P.~Kumar,
\newblock Phys. Rev. {\bf D80}, 083529 (2009), 0908.2430.

\bibitem{GomezReino:2006dk}
M.~Gomez-Reino and C.~A. Scrucca,
\newblock JHEP {\bf 05}, 015 (2006), hep-th/0602246.

\bibitem{GomezReino:2006wv}
M.~Gomez-Reino and C.~A. Scrucca,
\newblock JHEP {\bf 09}, 008 (2006), hep-th/0606273.

\bibitem{Dine:1981gu}
M.~Dine and W.~Fischler,
\newblock Phys. Lett. {\bf 110B}, 227 (1982).

\bibitem{Randall:1998uk}
L.~Randall and R.~Sundrum,
\newblock Nucl. Phys. {\bf B557}, 79 (1999), hep-th/9810155.

\bibitem{Giudice:1998xp}
G.~F. Giudice, M.~A. Luty, H.~Murayama, and R.~Rattazzi,
\newblock JHEP {\bf 12}, 027 (1998), hep-ph/9810442.

\bibitem{Gaillard:1999yb}
M.~K. Gaillard, B.~D. Nelson, and Y.-Y. Wu,
\newblock Phys. Lett. {\bf B459}, 549 (1999), hep-th/9905122.

\bibitem{Acharya:2008zi}
B.~S. Acharya, K.~Bobkov, G.~L. Kane, J.~Shao, and P.~Kumar,
\newblock Phys. Rev. {\bf D78}, 065038 (2008), 0801.0478.

\bibitem{Kaufman:2013pya}
B.~L. Kaufman, B.~D. Nelson, and M.~K. Gaillard,
\newblock Phys. Rev. {\bf D88}, 025003 (2013), 1303.6575.

\bibitem{Kaufman:2013oaa}
B.~Kaufman and B.~D. Nelson,
\newblock Phys. Rev. {\bf D89}, 085029 (2014), 1312.6621.

\bibitem{Everett:2015dqa}
L.~L. Everett, T.~Garon, B.~L. Kaufman, and B.~D. Nelson,
\newblock Phys. Rev. {\bf D93}, 055031 (2016), 1510.05692.

\bibitem{McDermott:2010pa}
S.~D. McDermott, H.-B. Yu, and K.~M. Zurek,
\newblock Phys. Rev. {\bf D83}, 063509 (2011), 1011.2907.

\bibitem{Fields:2006ga}
B.~Fields and S.~Sarkar,
\newblock (2006), astro-ph/0601514.

\bibitem{Talk:2015}
J.~Lesgourgues,
\newblock {Neutrino Cosmology from Planck 2014},
\newblock
  \url{http://www.cosmos.esa.int/documents/387566/387653/Ferrara_Dec4_11h50_Lesgourgues_NeutrinosReview.pdf}.

\bibitem{Hebecker:2014gka}
A.~Hebecker, P.~Mangat, F.~Rompineve, and L.~T. Witkowski,
\newblock JHEP {\bf 09}, 140 (2014), 1403.6810.

\bibitem{Allahverdi:2014ppa}
R.~Allahverdi, M.~Cicoli, B.~Dutta, and K.~Sinha,
\newblock JCAP {\bf 1410}, 002 (2014), 1401.4364.

\bibitem{Cicoli:2015bpq}
M.~Cicoli and F.~Muia,
\newblock JHEP {\bf 12}, 152 (2015), 1511.05447.

\bibitem{Acharya:2015zfk}
B.~S. Acharya and C.~Pongkitivanichkul,
\newblock (2015), 1512.07907.

\bibitem{Stott:2017hvl}
M.~J. Stott, D.~J.~E. Marsh, C.~Pongkitivanichkul, L.~C. Price, and B.~S.
  Acharya,
\newblock (2017), 1706.03236.

\bibitem{Abazajian:2016yjj}
CMB-S4, K.~N. Abazajian {\em et~al.},
\newblock (2016), 1610.02743.

\bibitem{Reece:2015lch}
M.~Reece and T.~Roxlo,
\newblock JHEP {\bf 09}, 096 (2016), 1511.06768.

\bibitem{Adshead:2016xxj}
P.~Adshead, Y.~Cui, and J.~Shelton,
\newblock JHEP {\bf 06}, 016 (2016), 1604.02458.

\bibitem{Tenkanen:2016jic}
T.~Tenkanen and V.~Vaskonen,
\newblock Phys. Rev. {\bf D94}, 083516 (2016), 1606.00192.

\bibitem{Hardy:2017wkr}
E.~Hardy and J.~Unwin,
\newblock (2017), 1703.07642.

\bibitem{Schabinger:2005ei}
R.~Schabinger and J.~D. Wells,
\newblock Phys. Rev. {\bf D72}, 093007 (2005), hep-ph/0509209.

\bibitem{Patt:2006fw}
B.~Patt and F.~Wilczek,
\newblock (2006), hep-ph/0605188.

\bibitem{MarchRussell:2008yu}
J.~March-Russell, S.~M. West, D.~Cumberbatch, and D.~Hooper,
\newblock JHEP {\bf 07}, 058 (2008), 0801.3440.

\bibitem{Ibanez:1991pr}
L.~E. Ibanez and G.~G. Ross,
\newblock Nucl. Phys. {\bf B368}, 3 (1992).

\bibitem{Dreiner:2005rd}
H.~K. Dreiner, C.~Luhn, and M.~Thormeier,
\newblock Phys. Rev. {\bf D73}, 075007 (2006), hep-ph/0512163.

\bibitem{Arvanitaki:2009hb}
A.~Arvanitaki, N.~Craig, S.~Dimopoulos, S.~Dubovsky, and J.~March-Russell,
\newblock Phys. Rev. {\bf D81}, 075018 (2010), 0909.5440.

\bibitem{Choi:2006fz}
S.~Y. Choi, H.~E. Haber, J.~Kalinowski, and P.~M. Zerwas,
\newblock Nucl. Phys. {\bf B778}, 85 (2007), hep-ph/0612218.

\bibitem{Babu:1997st}
K.~S. Babu, C.~F. Kolda, and J.~March-Russell,
\newblock Phys. Rev. {\bf D57}, 6788 (1998), hep-ph/9710441.

\bibitem{Feldman:2010wy}
D.~Feldman, Z.~Liu, P.~Nath, and G.~Peim,
\newblock Phys. Rev. {\bf D81}, 095017 (2010), 1004.0649.

\bibitem{Djouadi:2001fa}
A.~Djouadi, Y.~Mambrini, and M.~Muhlleitner,
\newblock Eur. Phys. J. {\bf C20}, 563 (2001), hep-ph/0104115.

\bibitem{Khachatryan:2016unx}
CMS, V.~Khachatryan {\em et~al.},
\newblock Phys. Rev. D  (2016), 1610.05133,
\newblock [Phys. Rev.D95,012009(2017)].

\bibitem{Aad:2015rba}
ATLAS, G.~Aad {\em et~al.},
\newblock Phys. Rev. {\bf D92}, 072004 (2015), 1504.05162.

\bibitem{Nelson:2002fk}
B.~D. Nelson,
\newblock {Anomaly Mediated SUSY Breaking From a String Theory Perspective},
\newblock in {\em {Proceedings, 1st String Phenomenology 2002 (StringPheno
  2002): Oxford, England, July 6-11, 2002}}, pp. 295--298, 2003,
  hep-ph/0211087.

\bibitem{Bullimore:2010aj}
M.~Bullimore, J.~P. Conlon, and L.~T. Witkowski,
\newblock JHEP {\bf 11}, 142 (2010), 1009.2380.

\bibitem{Goodsell:2010ie}
M.~Goodsell and A.~Ringwald,
\newblock Fortsch. Phys. {\bf 58}, 716 (2010), 1002.1840.

\bibitem{Goodsell:2011wn}
M.~Goodsell, S.~Ramos-Sanchez, and A.~Ringwald,
\newblock JHEP {\bf 01}, 021 (2012), 1110.6901.

\bibitem{Akerib:2016vxi}
LUX, D.~S. Akerib {\em et~al.},
\newblock Phys. Rev. Lett. {\bf 118}, 021303 (2017), 1608.07648.

\bibitem{Agnese:2015nto}
SuperCDMS, R.~Agnese {\em et~al.},
\newblock Phys. Rev. Lett. {\bf 116}, 071301 (2016), 1509.02448.

\bibitem{Angloher:2015ewa}
CRESST, G.~Angloher {\em et~al.},
\newblock Eur. Phys. J. {\bf C76}, 25 (2016), 1509.01515.

\bibitem{Akerib:2015cja}
LZ, D.~S. Akerib {\em et~al.},
\newblock (2015), 1509.02910.

\bibitem{Agnese:2016cpb}
SuperCDMS, R.~Agnese {\em et~al.},
\newblock Submitted to: Phys. Rev. D  (2016), 1610.00006.

\bibitem{Aalbers:2016jon}
DARWIN, J.~Aalbers {\em et~al.},
\newblock JCAP {\bf 1611}, 017 (2016), 1606.07001.

\bibitem{Frandsen:2012rk}
M.~T. Frandsen, F.~Kahlhoefer, A.~Preston, S.~Sarkar, and K.~Schmidt-Hoberg,
\newblock JHEP {\bf 07}, 123 (2012), 1204.3839.

\bibitem{Essig:2011nj}
R.~Essig, J.~Mardon, and T.~Volansky,
\newblock Phys. Rev. {\bf D85}, 076007 (2012), 1108.5383.

\bibitem{Graham:2012su}
P.~W. Graham, D.~E. Kaplan, S.~Rajendran, and M.~T. Walters,
\newblock Phys. Dark Univ. {\bf 1}, 32 (2012), 1203.2531.

\bibitem{Essig:2012yx}
R.~Essig, A.~Manalaysay, J.~Mardon, P.~Sorensen, and T.~Volansky,
\newblock Phys. Rev. Lett. {\bf 109}, 021301 (2012), 1206.2644.

\bibitem{Agnese:2013jaa}
SuperCDMS, R.~Agnese {\em et~al.},
\newblock Phys. Rev. Lett. {\bf 112}, 041302 (2014), 1309.3259.

\bibitem{Essig:2015cda}
R.~Essig {\em et~al.},
\newblock JHEP {\bf 05}, 046 (2016), 1509.01598.

\bibitem{Lee:2015qva}
S.~K. Lee, M.~Lisanti, S.~Mishra-Sharma, and B.~R. Safdi,
\newblock Phys. Rev. {\bf D92}, 083517 (2015), 1508.07361.

\bibitem{Hochberg:2015pha}
Y.~Hochberg, Y.~Zhao, and K.~M. Zurek,
\newblock Phys. Rev. Lett. {\bf 116}, 011301 (2016), 1504.07237.

\bibitem{Hochberg:2015fth}
Y.~Hochberg, M.~Pyle, Y.~Zhao, and K.~M. Zurek,
\newblock JHEP {\bf 08}, 057 (2016), 1512.04533.

\bibitem{Hochberg:2016ntt}
Y.~Hochberg, Y.~Kahn, M.~Lisanti, C.~G. Tully, and K.~M. Zurek,
\newblock (2016), 1606.08849.

\bibitem{Alexander:2016aln}
J.~Alexander {\em et~al.},
\newblock {Dark Sectors 2016 Workshop: Community Report},
\newblock 2016, 1608.08632.

\bibitem{Essig:2017kqs}
R.~Essig, T.~Volansky, and T.-T. Yu,
\newblock (2017), 1703.00910.

\bibitem{Curtin:2014cca}
D.~Curtin, R.~Essig, S.~Gori, and J.~Shelton,
\newblock JHEP {\bf 02}, 157 (2015), 1412.0018.

\bibitem{Lees:2014xha}
BaBar, J.~P. Lees {\em et~al.},
\newblock Phys. Rev. Lett. {\bf 113}, 201801 (2014), 1406.2980.

\bibitem{deNiverville:2011it}
P.~deNiverville, M.~Pospelov, and A.~Ritz,
\newblock Phys. Rev. {\bf D84}, 075020 (2011), 1107.4580.

\bibitem{Batell:2014mga}
B.~Batell, R.~Essig, and Z.~Surujon,
\newblock Phys. Rev. Lett. {\bf 113}, 171802 (2014), 1406.2698.

\bibitem{Essig:2016crl}
R.~Essig, J.~Mardon, O.~Slone, and T.~Volansky,
\newblock Phys. Rev. {\bf D95}, 056011 (2017), 1608.02940.

\bibitem{Billard:2013qya}
J.~Billard, L.~Strigari, and E.~Figueroa-Feliciano,
\newblock Phys. Rev. {\bf D89}, 023524 (2014), 1307.5458.

\bibitem{Svrcek:2006yi}
P.~Svrcek and E.~Witten,
\newblock JHEP {\bf 06}, 051 (2006), hep-th/0605206.

\bibitem{Arvanitaki:2009fg}
A.~Arvanitaki, S.~Dimopoulos, S.~Dubovsky, N.~Kaloper, and J.~March-Russell,
\newblock Phys. Rev. {\bf D81}, 123530 (2010), 0905.4720.

\bibitem{Acharya:2000gb}
B.~S. Acharya,
\newblock (2000), hep-th/0011089.

\bibitem{Witten:2001bf}
E.~Witten,
\newblock {Deconstruction, G(2) holonomy, and doublet triplet splitting},
\newblock in {\em {Supersymmetry and unification of fundamental interactions.
  Proceedings, 10th International Conference, SUSY'02, Hamburg, Germany, June
  17-23, 2002}}, pp. 472--491, 2001, hep-ph/0201018.

\bibitem{Friedmann:2002ty}
T.~Friedmann and E.~Witten,
\newblock Adv. Theor. Math. Phys. {\bf 7}, 577 (2003), hep-th/0211269.

\end{thebibliography}

\end{document}